\documentclass[twocolumn]{aastex631}

\usepackage{amsmath}

\begin{document}

\title[Spitzer 4.5 $\mu$m Phase Curve Survey]{A Comprehensive Analysis Spitzer 4.5 $\mu$m Phase Curve of Hot Jupiters}

\correspondingauthor{Lisa Dang}
\email{kha.han.lisa.dang@umontreal.ca}

\author[0000-0003-4987-6591]{Lisa Dang}
\affiliation{Trottier Institute for Research on Exoplanets and D\'epartement de Physique, Université de Montréal, 1375 Avenue Thérèse-Lavoie-Roux, Montréal, QC, H2V 0B3, Canada}

\author[0000-0003-4177-2149]{Taylor J. Bell}
\affiliation{Bay Area Environmental Research Institute, NASA Ames Research Center, Moffett Field, CA 94035}
\affiliation{Space Science and Astrobiology Division, NASA Ames Research Center Moffett Field, CA 94035}

\author[0009-0002-5701-6276]{Ying (Zoe) Shu}
\affiliation{Trottier Institute for Research on Exoplanets and D\'epartement de Physique, Université de Montréal, 1375 Avenue Thérèse-Lavoie-Roux, Montréal, QC, H2V 0B3, Canada}

\author[0000-0001-6129-5699]{Nicolas B. Cowan}
\affiliation{Department of Physics, McGill University, 3600 University St, Montreal, QC H3A 2T8, Canada}
\affiliation{Department of Earth \& Planetary Sciences, McGill University, 3450 University St, Montréal, H3A 2A7, Canada}

\author[0000-0003-4733-6532]{Jacob L.\ Bean}
\affiliation{Department of Astronomy and Astrophysics, University of Chicago, Chicago, IL 60637, USA}

\author[0000-0001-5727-4094]{Drake Deming}
\affiliation{Department of Astronomy, University of Maryland, College Park, MD 20742-2421, USA}

\author[0000-0002-1337-9051]{Eliza M.-R. Kempton}
\affiliation{Department of Astronomy, University of Maryland, College Park, MD 20742-2421, USA}

\author[0000-0003-4241-7413]{Megan Weiner Mansfield}
\affiliation{Steward Observatory, University of Arizona, Tucson, AZ 85721, USA}
\affiliation{NHFP Sagan Fellow}

\author[0000-0003-3963-9672]{Emily Rauscher}
\affiliation{Department of Astronomy, University of Michigan, 1085 S. University, Ann Arbor, MI 48109}

\author[0000-0001-9521-6258]{Vivien Parmentier}
\affiliation{Laboratoire Lagrange, Observatoire de la Côte d’Azur, CNRS, Université Côte d’Azur, Nice, France}

\author[0000-0002-7352-7941]{Kevin B. Stevenson}
\affiliation{JHU Applied Physics Laboratory, 11100 Johns Hopkins Road, Laurel, MD 20723, USA}

\author{Mark Swain}
\affiliation{NASA Jet Propulsion Laboratory, 4800 Oak Grove Drive, Pasadena, CA 91109, USA}

\author[0000-0000-0000-0000]{Laura Kreidberg}
\affiliation{Max-Planck-Institut für Astronomie, Königstuhl 17, D-69117 Heidelberg, Germany}

\author[0000-0003-3759-9080]{Tiffany Kataria}
\affiliation{NASA Jet Propulsion Laboratory, 4800 Oak Grove Drive, Pasadena, CA 91109, USA}

\author{Jean-Michel D\'esert}
\affiliation{Anton Pannekoek Institute for Astronomy, University of Amsterdam, Noord Holland, NL-1090GE Amsterdam, the Netherlands}

\author{Robert Zellem}
\affiliation{NASA Goddard Space Flight Center, 8800 Greenbelt Road
Greenbelt, MD 20771, USA}

\author{Jonathan J. Fortney}
\affiliation{Department of Astronomy and Astrophysics, University of California, Santa Cruz, CA 95064, USA}

\author{Nikole K. Lewis}
\affiliation{Department of Astronomy and Carl Sagan Institute, Cornell University, 122 Sciences Drive, Ithaca, NY 14853, USA}

\author{Michael Line}
\affiliation{School of Earth and Space Exploration, Arizona State University, Tempe, AZ 85287, USA}

\author{Caroline Morley}
\affiliation{Department of Astronomy, University of Texas at Austin, 2515 Speedway, Austin, TX 78712}

\author{Adam Showman}
\affiliation{Lunar and Planetary Lab, University of Arizona Tucson, AZ 85721-0092, USA}

\begin{abstract}
Although exoplanetary science was not initially projected to be a substantial part of the Spitzer mission, its exoplanet observations set the stage for current and future surveys with JWST and Ariel. We present a comprehensive reduction and analysis of Spitzer's 4.5 $\mu$m phase curves of 29 hot Jupiters on low-eccentricity orbits. The analysis, performed with the Spitzer Phase Curve Analysis (SPCA) pipeline, confirms that BLISS mapping is the best detrending scheme for most, but not all, observations. Visual inspection remains necessary to ensure consistency across detrending methods due to the diversity of phase curve data and systematics. Regardless of the model selection scheme—whether using the lowest-BIC or a uniform detrending approach—we observe the same trends, or lack thereof. We explore phase curve trends as a function of irradiation temperature, orbital period, planetary radius, mass, and stellar effective temperature. We discuss the trends that are robustly detected and provide potential explanations for those that are not observed. While it is almost tautological that planets receiving greater instellation are hotter, we are still far from confirming dynamical theories of heat transport in hot Jupiter atmospheres due to the sample's diversity. Even among planets with similar temperatures, other factors like rotation and metallicity vary significantly. Larger, curated sample sizes and higher-fidelity phase curve measurements from JWST and Ariel are needed to firmly establish the parameters governing day–night heat transport on synchronously rotating planets.
\end{abstract}


\section{Introduction} \label{sec:intro}

Among the 4,700 identified transiting exoplanets, over a hundred hot Jupiters have had their atmospheres characterized with the Spitzer Space Telescope and the Hubble Space Telescope over the past decade. Most studies of hot Jupiter atmospheres have focused on individual planets rather than broader trends. However, population-level investigations are crucial for understanding the processes that determine the climates and observed spectra of these highly irradiated worlds.

Population-level studies of exoplanet atmospheres have provided insights into the dominant composition and thermal structures of hot Jupiter atmospheres using observations of eclipses \citep[][]{2011ApJ...729...54C, 2019AJ....158..217W, 2020AJ....159..137G, 2021A&A...648A.127B, 2021AJ....162...36W, 2021ApJ...923..242G, 2021NatAs...5.1224M, 2022ApJS..260....3C, 2023AJ....165..104D, 2023ApJS..269...31E, 2024ApJ...971...33W} and transits \citep[][]{2016Natur.529...59S, 2017ApJ...847L..22F, 2018AJ....155..156T, 2019MNRAS.482.1485P, 2022arXiv221100649E}. In particular, spectroscopic eclipse observations probe the dayside atmosphere of a planet, while transit spectroscopy probes the planet's upper atmosphere near the day--night terminator. While they constrain complementary regions of the atmosphere, these observations do not fully capture the inhomogeneous nature of hot Jupiter atmospheres, nor their dynamics.

A long-standing mystery is what parameters control the horizontal transport of energy in the atmospheres of short-period planets \citep{2018haex.bookE.116P, 2020SSRv..216..139S}. Full-orbit phase curves offer the most comprehensive view of a planet’s atmosphere at all longitudes. In particular, infrared phase curves record the global thermal emission, thereby placing constraints on the planet’s energy budget. They provide insights on atmospheric dynamics via the day--night temperature contrast and longitudinal shift of the hot spot. Although many observatories can perform phase curve measurements, the Spitzer Space Telescope was particularly well-suited for measuring thermal phase curves due to its then unique infrared access and uninterrupted continuous observations \citep{2020NatAs...4..453D}. One of Spitzer’s greatest legacies is a large sample of phase curve observations, comprising more than 60 datasets in the 3.6 and 4.5 $\mu$m channels.

Theory predicts that horizontal energy transport is controlled, to first order, by incident stellar flux, which we quantify via a planet's irradiation temperature, $T_{\rm irr} \equiv T_\star \sqrt{R_\star/a}$, where $a$ is the orbital semi-major axis, $R_{\star}$ is the stellar radius, and $T_{\star}$ is the stellar effective temperature. The irradiation temperature is the zero-albedo radiative equilibrium temperature at the sub-stellar point. It is proportional to the global equilibrium temperature, $T_{\rm eq} = (1/4)^{1/4} T_{\rm irr} \approx 0.7\,T_{\rm irr}$, a useful quantity for planets with modest horizontal temperature contrasts (i.e., okay for hot Jupiters, not for ultra-hot Jupiters).

On synchronously rotating gas giants, the transfer of energy from the permanently illuminated hemisphere to the dark side should be proportional to the ratio of radiative and advective timescales \citep[e.g.,][]{2011ApJ...729...54C, 2016ApJ...821...16K}. To first order, greater levels of irradiation produce higher temperatures, hence much shorter radiative timescales and only slightly shorter advective timescales, leading to decreased day-to-night heat transport, and consequently a larger phase curve amplitude and smaller hotspot offset. However, this relationship with phase curve observables is more complex as different components of atmospheric circulation can influence these outcomes independently. For example, drag timescales and the interplay between various circulation components \citep{2021PNAS..11822705H, 2024MNRAS.531.1056R} can result in phase curve amplitude and hotspot offset being controlled by different processes.

\begin{figure}[!htbp]
	\includegraphics[width=\linewidth]{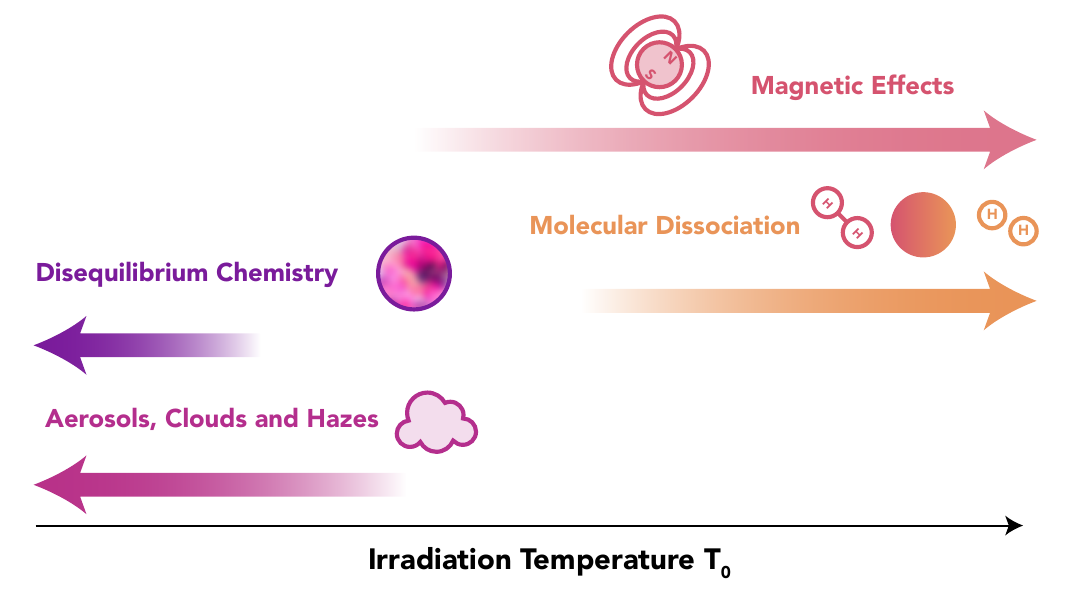}
    \includegraphics[width=\linewidth]{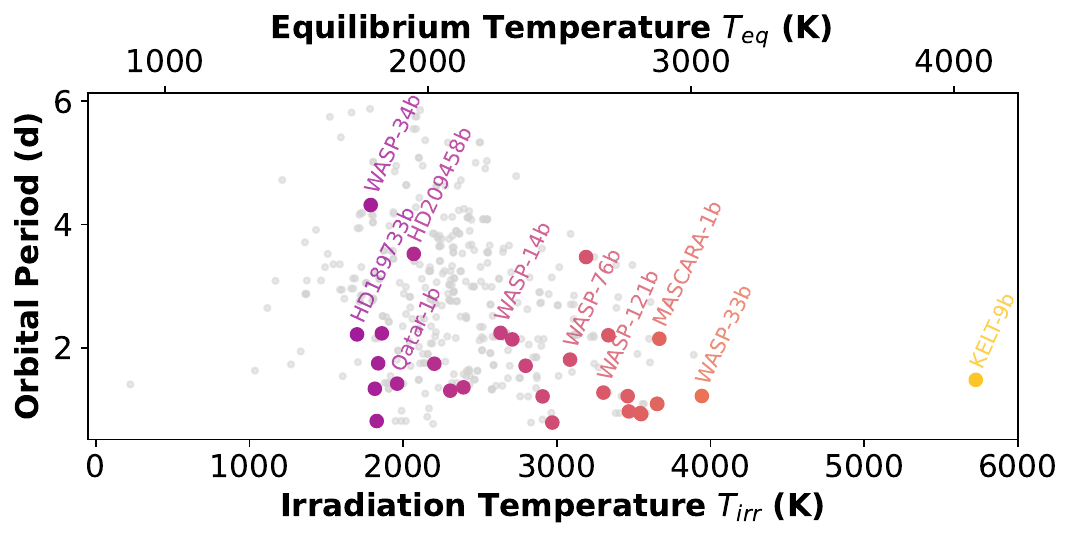}
    \vspace{-5mm}
    \caption{Schematic view of second-order effects beyond radiative forcing that could complicate the interpretation of phase curves for close-in exoplanets. These multi-factor effects motivate the large sample of phase curves used in this population study, with irradiation temperatures, equilibrium temperatures and orbital periods shown in the bottom panel. The grey points represent the known hot Jupiter population obtained from the NASA Exoplanet Archive, while the colored points represent our sample targets.}
\label{fig:Spitzer sample}
\end{figure}

Secondary factors beyond the radiative timescale can also affect the global atmospheric circulation on hot Jupiters, as depicted in Figure \ref{fig:Spitzer sample}. At high temperatures ($T_{\rm irr} > 2100$ K), hot Jupiters could experience changes in global circulation due to interactions between their partially ionized atmospheres and underlying magnetic fields. This could result in magnetic drag limiting the efficiency of heat recirculation to the nightside \citep[e.g.,][]{2012ApJ...751...59P, 2022AJ....163...35B} or introducing time-variable effects on the global-scale circulation \citep[e.g.,][]{2017NatAs...1E.131R}. Even at temperatures as low as $T_{\rm irr} = 1400$ K, thermo-resistive instability can produce significant variability \citep{2023arXiv230800892H}.

At even higher temperatures ($T_{\rm irr} > 3500$K), H$_{\rm 2}$ molecules can dissociate on the dayside of an ultra-hot Jupiter and then recombine on its nightside. Since dissociation takes energy and recombination releases it into the local atmosphere, these processes can enhance horizontal heat transport and reduce the day--night contrast \citep{2019MNRAS.489.1995B, 2019ApJ...886...26T, 2020ApJ...888L..15M}.

At lower temperatures, meanwhile, clouds and hazes can form and impact phase curve observables as well as the atmospheric temperature structure and dynamics \citep{2016MNRAS.455.2018D, 2016ApJ...828...22P, 2021ApJ...908..101R, 2023ApJ...951..117S}. Finally, at even lower temperatures, the chemical atmospheric state of the planet could be in disequilibrium, making the phase curve observables complicated to interpret \citep{2019ApJ...880...14S, 2023MNRAS.519.3129Z}. Fortunately, most Spitzer phase curves are of planets hotter than 1000 K, for which thermochemical equilibrium is a safe assumption \citep[but not perfect, as demonstrated by the SO$_2$ on the 1100 K WASP-39b;][]{2023Natur.617..483T}.

Recent comparative studies of Spitzer phase curves have revealed trends and variations in the longitudinal thermal maps and global atmospheric properties of highly irradiated gas giants \citep{2015MNRAS.449.4192S, 2017ApJ...850..154S, 2018AJ....155...83Z, 2019AJ....158..166B, 2019NatAs...3.1092K, 2021MNRAS.504.3316B, 2022AJ....163..256M}. 
A notable trend observed in these population studies is uniform nightside temperatures around $\sim$1000 K across various planets, regardless of stellar irradiation, while the dayside temperature increases with incident stellar flux. This uniformity suggests the presence of nightside clouds that block outgoing longwave radiation \citep{2019AJ....158..166B, 2019NatAs...3.1092K}.

Phase curve surveys have also reported tentative correlations between thermal phase curve properties and planetary system properties, e.g., stellar irradiation and orbital period, suggesting that a planet's orbital architecture may influence heat redistribution \citep{2018AJ....155...83Z, 2021MNRAS.504.3316B, 2022AJ....163..256M}. Thus, comparative studies of thermal eclipses---and ideally thermal phase curves---have provided valuable insights into how energy is transported from the perpetually lit daysides to the permanently obscured nightsides of these synchronously rotating planets.

In this paper, we present a comprehensive analysis of 4.5 $\mu$m phase curves of 29 hot Jupiters on circular orbits using the Spitzer Phase Curve Analysis Pipeline \citep[\texttt{SPCA};][]{2018NatAs...2..220D, 2021MNRAS.504.3316B}, where we experiment with multiple decorrelation methods to test the robustness of our retrieved trends. We describe our observations, reduction, and analysis in Section \ref{sec:methods}. In Section \ref{sec:results}, we discuss the results of our analysis and present empirical trends from our survey. We conclude in Section \ref{sec:conclusion}.

\section{Methods} \label{sec:methods}
\begin{figure*}
    \centering
	\includegraphics[width=0.93\linewidth]{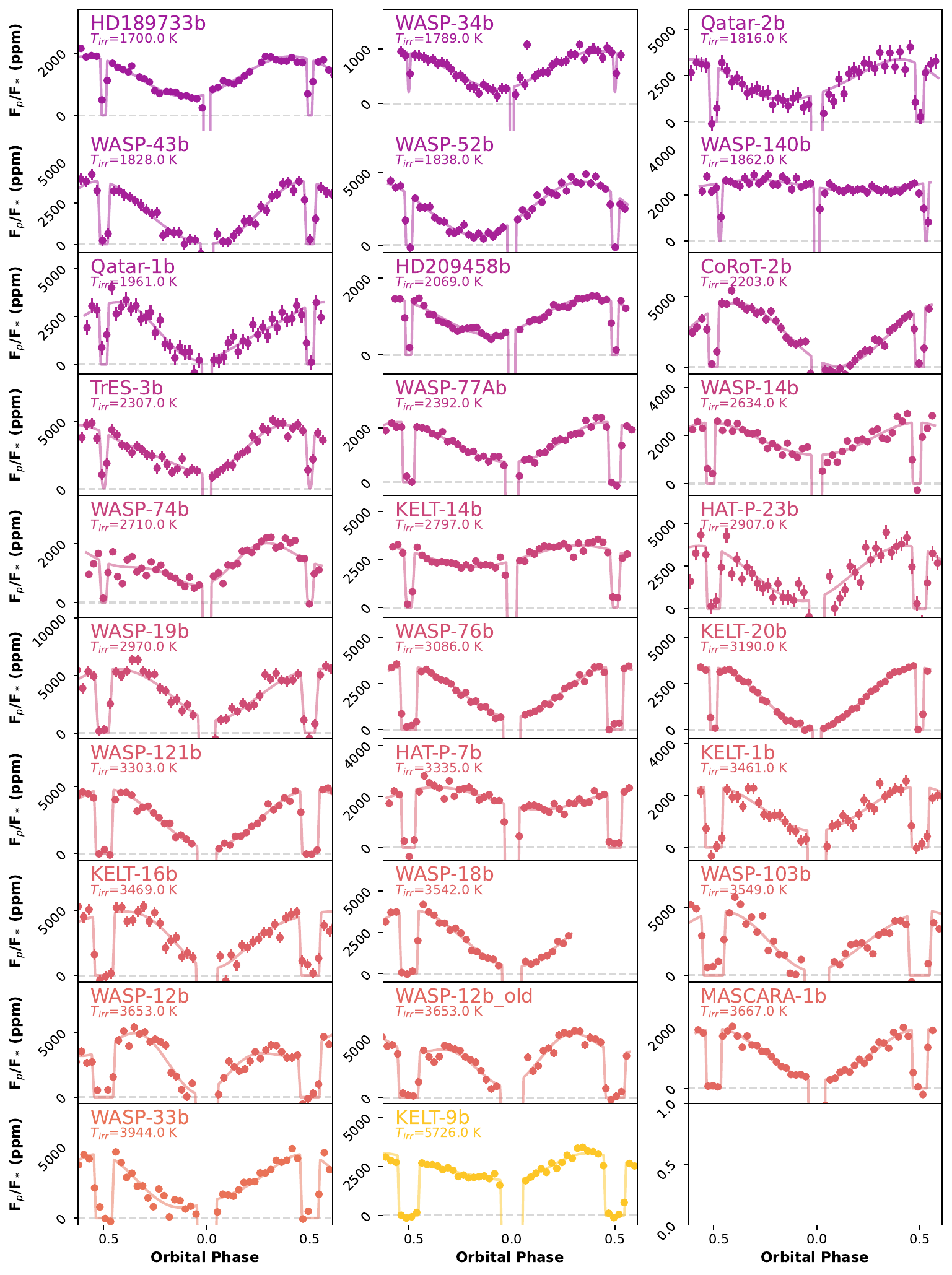}
    \vspace{-0mm}
    \caption{Preferred Detrended 4.5 $\mu$m Spitzer Phase Curves of hot Jupiters on circular orbits ordered by increasing levels of irradiation temperatures. The line represent the best astrophysical model and observations are binned for a total of 50 bins and the errorbars represent the scatter of each bin.}
    \label{fig:phasecurve_survey}
\end{figure*}

\begin{deluxetable}{lc}[hpbt!]
\renewcommand{\arraystretch}{1.0}
\tablecaption{Published Spitzer Phase Curves of Hot Jupiters on Circular Orbits}
\tablehead{
\colhead{Planet Name} & \colhead{Citation}
}
\startdata
Qatar-1b & \citet{2020AJ....159..225K} \\
KELT-9b & \citet{2020ApJ...888L..15M} \\
KELT-16b & \citet{2021MNRAS.504.3316B} \\
MASCARA-1b & \citet{2021MNRAS.504.3316B} \\
WASP-76b & \citet{2021AJ....162..158M} \\
Qatar-2b & \citet{2022AJ....163..256M} \\
WASP-52b & \citet{2022AJ....163..256M} \\
WASP-34b & \citet{2022AJ....163..256M} \\
WASP-140b & \citet{2022AJ....163..256M} \\
CoRoT-2b & \citet[][PID 11073]{2018NatAs...2..220D} \\
HAT-P-7b & \citet[][PID 60021]{2016ApJ...823..122W} \\
HD 189733b & \citet[][PID 60021]{2012ApJ...754...22K} \\
HD 209458b & \citet[][PID 60021]{2014ApJ...790...53Z} \\
KELT-1b & \citet[][PID 11095]{2019AJ....158..166B} \\
WASP-12b & \citet[][PID 70060]{2012ApJ...747...82C} \\
 & \citet[][PID 90186]{2019MNRAS.489.1995B}\\
WASP-14b & \citet[][PID 80073]{2015ApJ...811..122W} \\
WASP-18b & \citet[][PID 60185]{2013MNRAS.428.2645M} \\
WASP-19b & \citet[][PID 80073]{2016ApJ...823..122W} \\
WASP-33b & \citet[][PID 80073]{2018AJ....155...83Z} \\
WASP-43b & \citet[][PID 11001]{2017AJ....153...68S} \\
& \citet{2020AJ....160..140M} \\
WASP-103b & \citet[][PID 11099]{2018AJ....156...17K} \\
WASP-121b & \citet[][PID 13242]{2023morello}
\enddata
\tablecomments{For planets where no PID is provided the data were obtained as part of the Ultimate Spitzer Phase Curve Survey program (PID 13038, PI Stevenson; and PID 14059, PI Bean).}
\label{tab: published phase curves}
\end{deluxetable}

\subsection{Observations}\label{sec:observations}
This population study uses 4.5 $\mu$m full-orbit phase curves of hot Jupiters on circular orbits (with $e \leq 0.15$) observed with channel 2 of Spitzer's InfraRed Array Camera \citep[IRAC;][]{2004ApJS..154...10F, 2004ApJS..154....1W}. The targets in this sample are shown in Figure \ref{fig:Spitzer sample} and include phase curves from the Ultimate Spitzer Phase Curve Survey program (PID 13038, PI Stevenson; and PID 14059, PI Bean) as well as other hot Jupiter phase curves from smaller, targeted proposals (Table \ref{tab: published phase curves}). In particular, we present the first analysis of phase curve observations of WASP-74b \citep{2015AJ....150...18H}, KELT-20b \citep{2017AJ....154..194L}, TrES-3b \citep{2007ApJ...663L..37O}, WASP-77Ab \citep{2013PASP..125...48M}, KELT-14b \citep{2016AJ....151..138R}, and HAT-P-23b \citep{2011ApJ...742..116B}, along with a re-analysis of the published phase curves listed in Table \ref{tab: published phase curves}.

We exclude the phase curves of HD 149026b \citep[PID 60021][]{2018AJ....155...83Z}, WASP-95b, and KELT-7b (PID 14059) as our attempts to fit these datasets were unsuccessful. For this survey, we have also excluded the phase curves of the hot Neptune LTT 9779b \citep{2020ApJ...903L...7C} and ultra-short-period rocky planets (USPs) 55 Cnc e \citep{2016Natur.532..207D, 2022AJ....164..204M}, K2-141b \citep{2022A&A...664A..79Z}, and LHS 3844b \citep{2019Natur.573...87K} to maintain a uniform sample of hot Jupiters.

Each phase curve observation, except for WASP-18b, continuously spans the full orbit and typically begins before a secondary eclipse and ends after the subsequent secondary eclipse to better disentangle planetary variation from instrumental variation. Most observing datasets, except for WASP-103b, used the subarray mode and consist of 64-frame datacubes made of 32$\times$32 pixel images (corresponding to 39 arcsec$\times$39 arcsec). Details regarding the exposure durations and other observational parameters for each previously published dataset can be found within the corresponding papers.

\subsection{Comprehensive SPCA Analysis}
Spitzer IRAC observations are known to be affected by instrumental variations due to the non-uniform intra-pixel sensitivity of the detector and slight fluctuations of the telescope pointing \citep{2016AJ....152...44I}. A variety of decorrelation methods have been proposed over the years to detrend these instrumental variations. Since each phase curve dataset has its own quirks, there is no single decorrelation method that is best suited for all observations. We analyze each Spitzer dataset using the Spitzer Phase Curve Analysis Pipeline \citep[\texttt{SPCA};][]{2018NatAs...2..220D, 2021MNRAS.504.3316B}. \texttt{SPCA} is an open-source pipeline for the analysis of Spitzer/IRAC channel 1 and 2 time-series photometry, incorporating some of the most commonly used decorrelation methods. We first perform a pixel-level 5$\sigma$-clip along the entire time-series and mask any outliers and other hot pixels. We also apply a frame-by-frame background subtraction by subtracting the median value of the pixels outside of an $8 \times 8$ central pixel box. We then extract the raw time-series photometry by experimenting with point-spread-function (PSF) fitting and aperture photometry with a variety of aperture sizes and shapes. At this step, we also determine the target's centroid ($x, y$) and the target's PSF width along each axis ($\sigma _x, \sigma _y$) in each frame with a flux-weighted mean (FWM) routine on the central $5 \times 5$ pixels. We then select the photometry with the lowest root-mean-square scatter for the rest of the analysis as detailed in \citet{2021MNRAS.504.3316B}. In addition, we also carry out a different photometry routine by extracting each pixel's lightcurve within a $3 \times 3$ or $5 \times 5$ box centered on the pixel position (15, 15) in order to apply Pixel Level Decorrelation \citep[PLD;][]{2015ApJ...805..132D}. We then perform a final 5$\sigma$-clip along the entire time-series to discard measurements with outlying flux or centroids.

We then decorrelate each raw phase curve by simultaneously fitting an astrophysical phase curve model and a multiplicative instrumental detector noise model. For the astrophysical phase curve model, we use \texttt{batman} \citep{2015PASP..127.1161K} to model the transit and secondary eclipse, and we experiment with a first- and second-order Fourier series to model the phase variation as described in \citet{2021MNRAS.504.3316B}. We impose a Gaussian prior, using the most precise constraints from the NASA Exoplanet Archive\footnote{\url{https://exoplanetarchive.ipac.caltech.edu/}} by using the \texttt{exofile} package\footnote{\url{https://github.com/AntoineDarveau/exofile}} to extract precise constraints for each orbital parameter, such as the time of transit, $t_0$, the orbital period, $P$, the semi-major axis, $a/R_*$, and the orbital inclination, $i$, as our single full-orbit phase curves are unlikely to provide better constraints than the long-baseline observations used to discover these planets. We also place uniform priors on the planet and stellar radius ratio $R_p/R_*$ and eclipse depth $F_p/F_*$ to allow them to vary between 0 and 1, as this minimally informative prior spans the full range of physically plausible values. Lastly, we adopt a prior on the phase curve coefficients to ensure that the phase curves are positive, i.e., not allowing the phase variation to dip below the stellar flux baseline. 

We use multiple intra-pixel sensitivity variation models that are implemented within \texttt{SPCA} such as 2D polynomials \citep{2005ApJ...626..523C}, BiLinearly-Interpolated Subpixel Sensitivity (BLISS) mapping \citep{2012ApJ...754..136S}, and Pixel Level Decorrelation \citep{2015ApJ...805..132D}. In particular, we test 2D polynomials of order 2 to 5 with the target's centroid as a covariate. For BLISS mapping, we follow the method described in \cite{2021MNRAS.504.3316B} to determine the optimal BLISS map resolution for each dataset since the distribution of centroids differs for each phase curve observation. We also experiment with a $3 \times 3$ and $5 \times 5$ pixel stamp for Pixel Level Decorrelation that uses the fractional flux measured by each pixel as a regressor for the detector model, and we also experiment with a first- and second-order PLD model \citep[e.g.,][]{2016AJ....152..100L, 2018AJ....155...83Z}.

\renewcommand{\arraystretch}{1.2}
\begin{deluxetable*}{lCCCCCCl}[t!]
\tablecaption{Preferred model parameters for each of our fitted phase curves to Spitzer 4.5 $\mu$m full-orbit phase observations \label{tab:bestfit_table}}
\tablewidth{0pt}
\tablehead{
\colhead{Planet} & \colhead{Mode} & \colhead{$R_p/R_{\star}$} & \colhead{$F_p/F_{\star}$ (ppm)} & \colhead{$F_{\rm min}$ (ppm)} & \colhead{$F_{\rm max}$ (ppm)} & \colhead{Offset ($^{\circ}$E)} & \colhead{$T_0$ (K)} 
}
\startdata
HD189733b & \rm Poly5 & $0.1564^{+0.0002} _{-0.0002}$ & $1800^{+20} _{-30}$ & $660^{+80} _{-70}$ & $1900^{+30} _{-30}$ & $34^{+3} _{-2}$ & 1700\\ 
WASP-34b & \rm BLISS & $0.127^{+0.004} _{-0.004}$ & $900^{+300} _{-200}$ & $300^{+300} _{-200}$ & $900^{+300} _{-200}$ & $20^{+10} _{-10}$ & 1789\\ 
Qatar-2b & \rm BLISS & $0.165^{+0.001} _{-0.001}$ & $3100^{+300} _{-300}$ & $1200^{+700} _{-700}$ & $3200^{+400} _{-400}$ & $30^{+20} _{-20}$ & 1816\\ 
WASP-43b & \rm PLDAper1-3x3 & $0.159^{+0.001} _{-0.001}$ & $3800^{+100} _{-100}$ & $200^{+200} _{-100}$ & $3900^{+100} _{-100}$ & $9^{+3} _{-3}$ & 1828\\ 
WASP-52b & \rm BLISS & $0.156^{+0.002} _{-0.002}$ & $3500^{+300} _{-300}$ & $800^{+700} _{-600}$ & $4300^{+500} _{-500}$ & $60^{+10} _{-20}$ & 1837\\ 
WASP-140b & \rm BLISS & $0.155^{+0.008} _{-0.005}$ & $2500^{+600} _{-300}$ & $2200^{+600} _{-500}$ & $3100^{+600} _{-400}$ & $-40^{+90} _{-40}$ & 1861\\ 
Qatar-1b & \rm BLISS & $0.146^{+0.002} _{-0.002}$ & $3100^{+300} _{-300}$ & $500^{+500} _{-300}$ & $3200^{+300} _{-300}$ & $-30^{+20} _{-20}$ & 1961\\ 
HD209458b & \rm Poly5 & $0.1205^{+0.0004} _{-0.0004}$ & $1380^{+50} _{-40}$ & $300^{+100} _{-100}$ & $1470^{+40} _{-40}$ & $43^{+5} _{-6}$ & 2069\\ 
CoRoT-2b & \rm Poly4 & $0.17^{+0.001} _{-0.002}$ & $4900^{+200} _{-200}$ & $50^{+70} _{-30}$ & $4400^{+200} _{-200}$ & $-39^{+3} _{-3}$ & 2203\\ 
TrES-3b & \rm Poly4 & $0.169^{+0.01} _{-0.005}$ & $4200^{+500} _{-400}$ & $1100^{+800} _{-700}$ & $4800^{+700} _{-500}$ & $50^{+10} _{-10}$ & 2306\\ 
WASP-77Ab & \rm BLISS & $0.109^{+0.0007} _{-0.0008}$ & $2180^{+90} _{-90}$ & $700^{+200} _{-200}$ & $2190^{+90} _{-100}$ & $14^{+8} _{-8}$ & 2392\\ 
WASP-14b & \rm Poly4 & $0.0946^{+0.0006} _{-0.0006}$ & $2540^{+80} _{-80}$ & $1180^{+80} _{-90}$ & $2550^{+80} _{-80}$ & $2^{+4} _{-4}$ & 2633\\ 
WASP-74b & \rm PLDAper1-5x5 & $0.099^{+0.002} _{-0.001}$ & $1380^{+90} _{-100}$ & $600^{+100} _{-100}$ & $2100^{+100} _{-100}$ & $72^{+3} _{-3}$ & 2710\\ 
KELT-14b & \rm BLISS & $0.112^{+0.002} _{-0.002}$ & $2900^{+200} _{-100}$ & $2100^{+300} _{-400}$ & $3100^{+200} _{-200}$ & $60^{+10} _{-20}$ & 2797\\ 
HAT-P-23b & \rm BLISS & $0.115^{+0.002} _{-0.002}$ & $3500^{+200} _{-200}$ & $400^{+400} _{-300}$ & $3700^{+300} _{-300}$ & $28^{+8} _{-9}$ & 2907\\ 
WASP-19b & \rm Poly3 & $0.139^{+0.002} _{-0.002}$ & $5500^{+300} _{-300}$ & $1200^{+400} _{-400}$ & $5700^{+300} _{-300}$ & $-17^{+4} _{-4}$ & 2970\\ 
WASP-76b & \rm Poly5 & $0.1049^{+0.0008} _{-0.0009}$ & $3400^{+100} _{-100}$ & $900^{+300} _{-300}$ & $3400^{+100} _{-100}$ & $-3^{+7} _{-9}$ & 3085\\ 
KELT-20b & \rm BLISS & $0.116^{+0.0007} _{-0.0007}$ & $3380^{+80} _{-70}$ & $100^{+100} _{-100}$ & $3410^{+80} _{-70}$ & $3^{+4} _{-3}$ & 3190\\ 
WASP-121b & \rm BLISS & $0.1249^{+0.0009} _{-0.0009}$ & $4800^{+100} _{-100}$ & $600^{+400} _{-300}$ & $4900^{+100} _{-100}$ & $-5^{+4} _{-4}$ & 3303\\ 
HAT-P-7b & \rm BLISS & $0.077^{+0.001} _{-0.001}$ & $2200^{+100} _{-100}$ & $1500^{+300} _{-300}$ & $2400^{+200} _{-200}$ & $-60^{+20} _{-20}$ & 3335\\ 
KELT-1b & \rm Poly4 & $0.076^{+0.001} _{-0.001}$ & $2300^{+100} _{-100}$ & $600^{+200} _{-200}$ & $2400^{+100} _{-100}$ & $5^{+5} _{-5}$ & 3461\\ 
KELT-16b & \rm Poly4 & $0.107^{+0.002} _{-0.002}$ & $4700^{+300} _{-300}$ & $500^{+600} _{-400}$ & $5000^{+300} _{-300}$ & $-10^{+30} _{-30}$ & 3468\\ 
WASP-18b & \rm PLDAper1-3x3 & $0.0977^{+0.0005} _{-0.0005}$ & $3900^{+100} _{-100}$ & $500^{+100} _{-100}$ & $3900^{+100} _{-100}$ & $-4^{+2} _{-2}$ & 3542\\ 
WASP-103b & \rm BLISS & $0.1155^{+0.0009} _{-0.001}$ & $5200^{+200} _{-100}$ & $300^{+200} _{-200}$ & $4700^{+100} _{-100}$ & $-15^{+4} _{-4}$ & 3548\\ 
WASP-12b (visit 1) & \rm Poly3 & $0.108^{+0.002} _{-0.001}$ & $4100^{+200} _{-200}$ & $300^{+300} _{-200}$ & $5000^{+200} _{-200}$ & $-8^{+5} _{-4}$ & 3653\\ 
WASP-12b (visit 2) & \rm BLISS & $0.105^{+0.001} _{-0.001}$ & $4200^{+200} _{-200}$ & $1000^{+400} _{-400}$ & $5300^{+300} _{-200}$ & $30^{+8} _{-8}$ & 3653\\ 
MASCARA-1b & \rm BLISS & $0.0788^{+0.0008} _{-0.0009}$ & $1950^{+80} _{-90}$ & $300^{+300} _{-200}$ & $1940^{+90} _{-90}$ & $-10^{+10} _{-10}$ & 3667\\ 
WASP-33b & \rm PLDAper1-3x3 & $0.1101^{+0.0005} _{-0.0005}$ & $4430^{+60} _{-60}$ & $770^{+50} _{-60}$ & $4510^{+50} _{-60}$ & $12^{+1} _{-1}$ & 3943\\ 
KELT-9b & \rm PLDAper2-3x3 & $0.0806^{+0.0006} _{-0.0005}$ & $2890^{+40} _{-40}$ & $1840^{+90} _{-110}$ & $3190^{+50} _{-50}$ & $49^{+4} _{-4}$ & 5726\\
\enddata
\tablecomments{In the table, uncertainties are rounded to one significant figures and the parameter values are rounded to match the precision of their associated errors.}
\end{deluxetable*}

\subsection{Model fit and Model Comparison}
For each model fit, we perform a Markov Chain Monte Carlo (MCMC) analysis using \texttt{emcee} \citep{2013PASP..125..306F}, where the astrophysical and instrumental models are fit simultaneously. Before the MCMC analysis, to ensure that our routine starts at a reasonable part of parameter space, we first fixed the astrophysical parameters to the most precise values found on the NASA Exoplanet Archive, and we performed a series of initial optimizations on the detector model parameters to ensure that our MCMC is initialized at a reasonable position. \texttt{SPCA} then performs additional optimization by exploring different parts of parameter space to find the highest log-likelihood location as the starting position of our MCMC, as detailed in \cite{2021MNRAS.504.3316B}. To initialize our MCMC marginalization, our 150 walkers are normally distributed around the maximum log-likelihood location with a standard deviation equivalent to 0.01\% of the parameter's value. We then run the MCMC for 5000 steps or until they meet our convergence criteria described in \cite{2018NatAs...2..220D}. Afterward, we run an additional 1000 MCMC steps and estimate each parameter's best-fit value and uncertainty by finding the maximum log-likelihood value and the 16th and 84th percentiles of each parameter's posterior distribution, respectively. 

To compare the different combinations of detector and astrophysical models, we calculate the Bayesian Information Criterion (BIC), defined as:
\begin{equation*}
    {\rm BIC} = -2\ln{L} + N_{\rm par} \ln{N_{\rm dat}},
\end{equation*}
\noindent where $L$ is the fit's log-likelihood, $N_{\rm par}$ is the number of free parameters in the fit, and $N_{\rm dat}$ is the number of data points. Some detector models can occasionally mimic low-frequency phase variations. To mitigate this, we perform a visual overview of all resulting fit combinations. For our nominal preferred model for each dataset, we first select the lowest-BIC model and visually inspect each phase curve's shape to ensure it is physically plausible, i.e., that the general shape of the phase curve does not look suspect and is also reproduced by other decorrelation methods (see section \ref{sec:SPCA example}). If the lowest-BIC phase curve shape is not reproduced by other decorrelation methods, we then look among the top-ranked models and select the phase curve that is generally reproduced by other methods. These preferred phase curve models are then used for our population-level analysis.

To ensure that our statistical trends are robust against our model selection criteria, we also perform two additional population-level analyses without visual inspection of the phase curve. In particular, we apply a uniform analysis using BLISS, as this is the decorrelation method that is generally preferred, and an analysis where we use the phase curves with the lowest BIC (see Appendix \ref{sec:robustness}). Though the individual phase curve shapes may vary between analyses, we find that our retrieved trends are generally consistent across our model selection criteria.

\subsection{Example of a SPCA Analysis Routine and Model Selection}\label{sec:SPCA example}
\begin{figure*}[h!]
    \centering
    \includegraphics[width=0.54\textwidth]{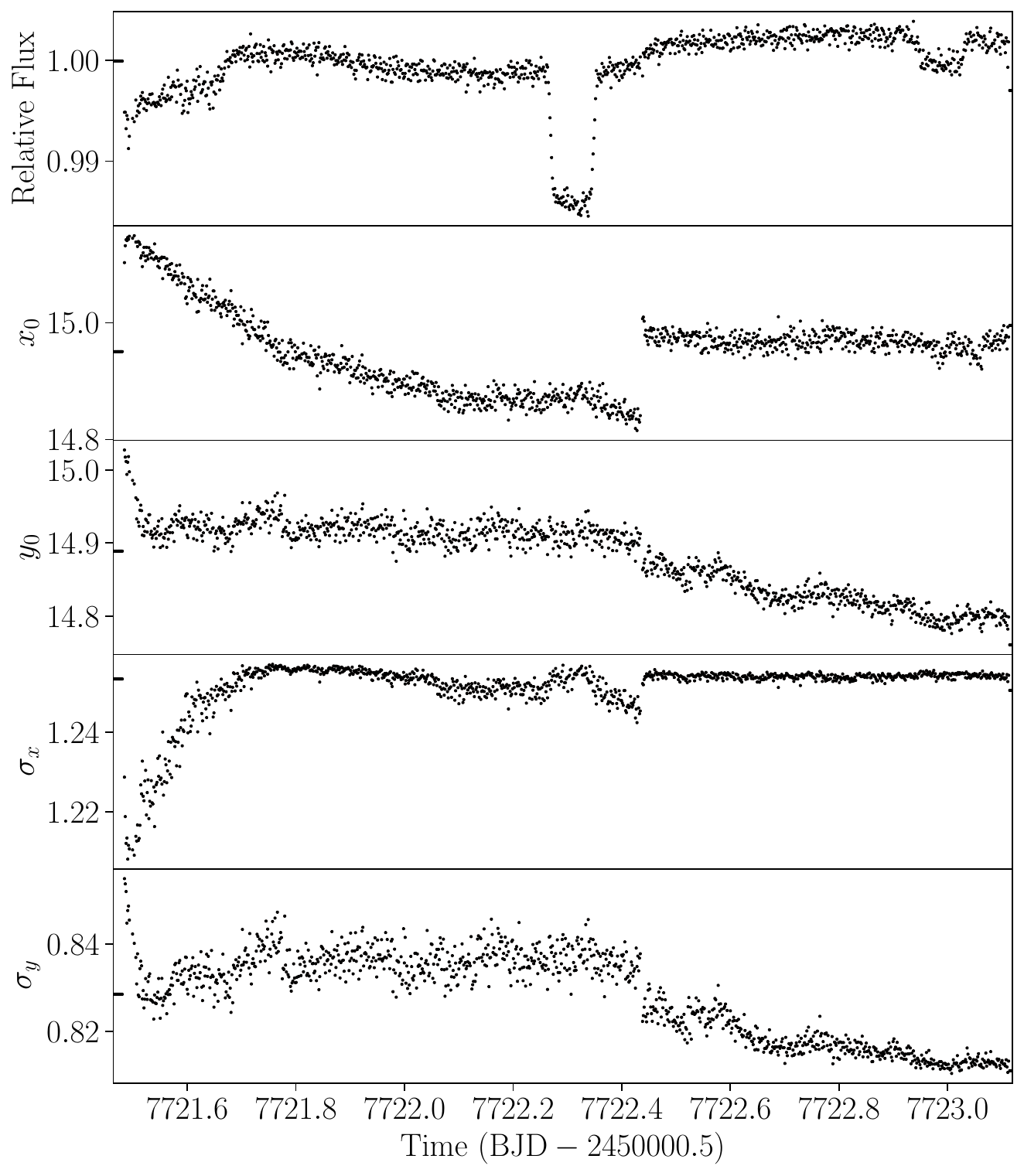}
    \includegraphics[width=0.4\textwidth]{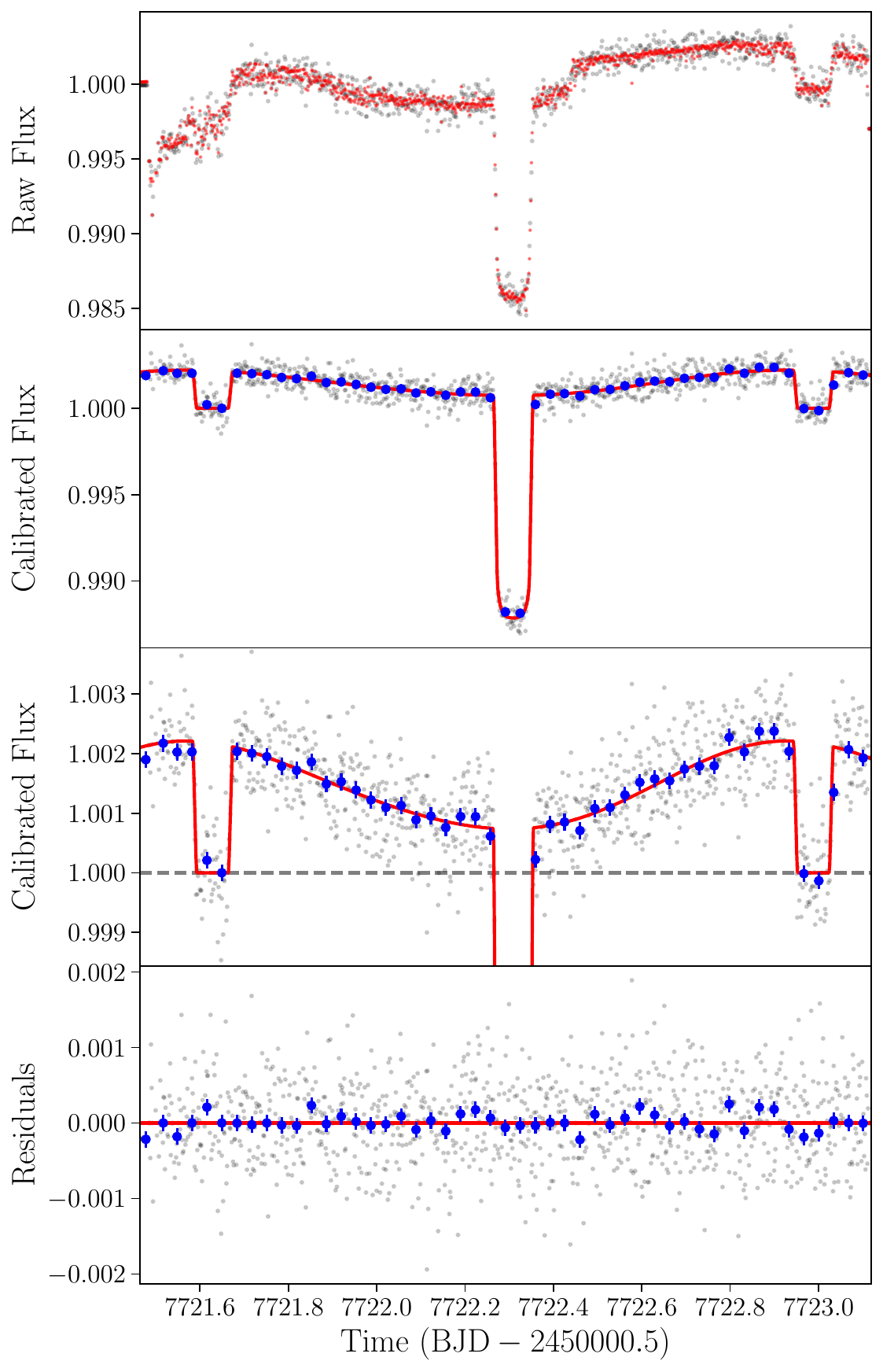}
    \includegraphics[width=\textwidth]{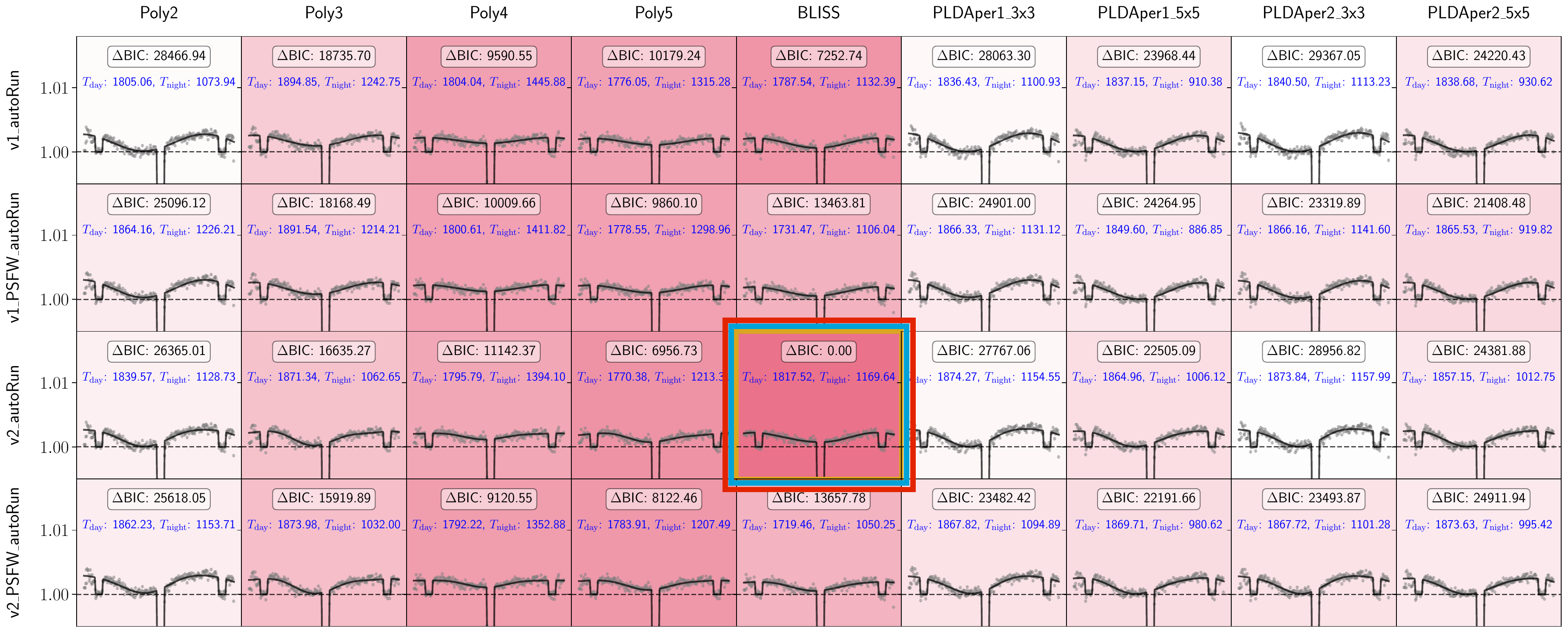}
    \caption{\textbf{Summary of the analysis of a Spitzer phase curve with SPCA} \emph{Top-left:} Example of the raw photometry output from SPCA. \emph{Top-right:} Preferred detrended phase curve of WASP-77Ab. \emph{Bottom:} A comprehensive overview of all SPCA fits of the 4.5 $\mu$m full-orbit phase curve of WASP-77Ab. Each subplot displays the best-fit astrophysical model compared to the corrected Spitzer photometry with different detector models. Each column indicates the detector model used. The rows indicate the phase variation model used and whether the detector model also included a linear model with PSF width ($\sigma_x$, $\sigma_y$) as inputs. The $\Delta$BIC from the preferred solution of each fit is indicated in each box, and the opacity of the background of each box reflects fit preference (darker is better). The dayside and nightside temperatures inferred from each fit are displayed.}
    \label{fig:SPCA_WASP-77Ab}
\end{figure*}

Figure \ref{fig:SPCA_WASP-77Ab} is an example of a comprehensive SPCA analysis for one data set and model selection process. First, the photometry and PSF properties are extracted from the time-series of 32$\times$32 pixel frames using multiple photometric schemes. The photometry resulting in the smallest root-mean-square (RMS) scatter is then selected and detrended by fitting different combinations of phase curve models and systematic noise models simultaneously. Finally, we use a mixture of BIC evaluation and visual inspection to determine the best-fit model for each phase observation.

\section{Results}\label{sec:results}
The preferred astrophysical models for each phase curve dataset are presented in Figure \ref{fig:phasecurve_survey}. The best-fit ratio of planet-to-star radius, $R_p/R_*$, eclipse depth $F_p/F_*$, and phase curve $F_{\rm max}$, $F_{\rm min}$, and offsets are reported in Table \ref{tab:bestfit_table}. For each phase curve, we convert dayside and nightside fluxes, i.e., the mid-eclipse flux and mid-transit flux \citep{2017ApJ...850..154S}, to brightness temperatures, $T_{\rm day}$ and $T_{\rm night}$, by inverting the Planck function. The phase curves are listed in order of increasing irradiation temperature. The values used to compute the irradiation temperature are listed in Table \ref{tab:exoparams_table} of Appendix \ref{sec:Appendix Target Params}.

\subsection{Bond Albedo and Heat Recirculation Efficiency}
\begin{figure}[t!]
	\includegraphics[width=\linewidth]{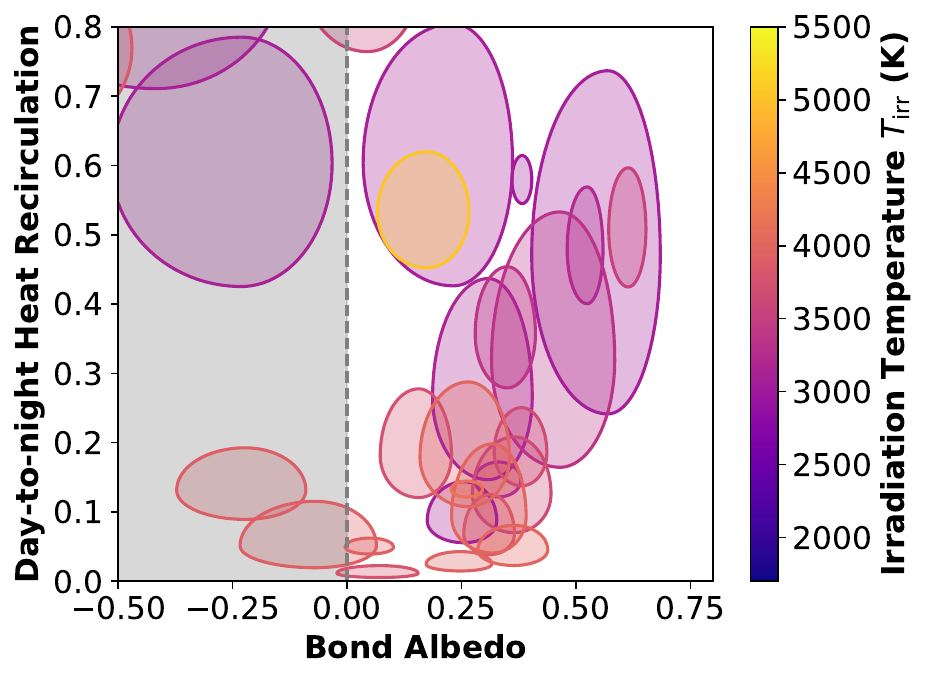}
 \vspace{-5mm}
    \caption{Inferred Bond albedo, $A_{\rm B}$, and heat recirculation efficiency, $\varepsilon$, from our sample of 4.5 $\mu$m Spitzer phase curves. Each contour represents the 1$\sigma$ confidence region estimates for each planet, and contour colors, corresponding to the colors in Figure 1, represent the irradiation temperature $T_{\rm irr}$.}
    \label{fig:BondAlb}
\end{figure}

Using the dayside and nightside brightness temperatures, we infer each planet's Bond albedo, $A_{\rm B}$, and recirculation efficiency, $\varepsilon$, by using the parametrization of \cite{2011ApJ...729...54C}:
\begin{equation}
\begin{aligned}
T_{\rm day} &= T_{\rm irr} (1 - A_{\rm B})^{1/4} \left( \frac{2}{3} - \frac{5}{12} \varepsilon \right)^{1/4}, \\
T_{\rm night} &= T_{\rm irr} (1 - A_{\rm B})^{1/4} \left( \frac{\varepsilon}{4} \right)^{1/4}.
\end{aligned}
\label{eq: bond-eps}
\end{equation}
\noindent Following \cite{2015MNRAS.449.4192S}, we generate a grid of $A_{\rm B}$ and $\varepsilon$ between 0 and 1, use Eq. (\ref{eq: bond-eps}) to predict dayside and nightside temperatures, and compare these to the measured $T_{\rm day}$ and $T_{\rm night}$ to create a $\chi^2$ surface for each planet. We then define the 1$\sigma$ confidence interval of $A_{\rm B}$ and $\varepsilon$, shown in Figure \ref{fig:BondAlb}, as the region where $\chi^2 \leq 1$ (also listed in Table \ref{tab:bestfit_table}).

\subsection{Statistical Significance of Trends}
We search for trends in phase curve properties against the following system parameters: the irradiation temperature $T_{\rm irr}$, the orbital period $P_{\rm orb}$, the planet's radius $R_p$, the stellar temperature $T_\star$, and the planet's mass $M_p$. The values for each planet are listed in Table \ref{tab:exoparams_table}.
We search for trends in $T_{\rm day}$ and $T_{\rm night}$, phase amplitude, phase offset, Bond albedo $A_{\rm B}$, and recirculation efficiency $\varepsilon$. We normalize the phase amplitude $(F_{\rm max} - F_{\rm min})/F_{\rm max}$ to facilitate comparison of systems with different orbital properties, and we elect to use the absolute phase offset to focus on the strength of energy transport rather than the direction of the energy transport.

In Figure \ref{fig:trends_figure}, we plot each of the phase curve properties against the independent quantities listed above. To account for the uncertainties on each phase curve property, we generate a Monte Carlo sampling centered on the estimated value with a standard deviation set to its uncertainty. For each scatter plot, we perform a linear regression to fit a line to the sample data points to estimate the linear correlation between the variables. We then calculate the Pearson correlation coefficient for each pair of variables to quantify the strength and direction of the linear relationship. The Pearson coefficient ranges from [-1, 1], with -1, 0, and 1 indicating a perfect negative relationship, no linear relationship, and a perfect positive relationship, respectively. The resulting coefficient is displayed on each panel of Figure \ref{fig:trends_figure}, and the background color of each subplot indicates the absolute value of the correlation coefficient, providing a quick visual indication of the strength of the correlation. We have also ensured that our analysis is robust against the model selection criteria chosen by performing the same analysis with the lowest-BIC sample and the BLISS-only sample. The resulting trends have similar slopes and dispersions.

\subsection{Trends in Phase Curve Measurements}

\begin{deluxetable}{lllll}
\tablecaption{Derived Parameters from Phase Curve Measurements \label{tab:params_pc_derived}}
\tablehead{
\colhead{Planet Name} & \colhead{T$_{\rm day}$ (K)} & \colhead{T$_{\rm night}$ (K)} & \colhead{A$_{\rm B}$} & \colhead{$\varepsilon$}
}
\startdata
HD189733b & $1217^{+6} _{-6}$ & $930^{+30} _{-30}$ & $0.38^{+0.03} _{-0.03}$ & $0.58^{+0.04} _{-0.04}$ \\
WASP-34b & $1200^{+100} _{-100}$ & $800^{+200} _{-200}$ & $0.6^{+0.1} _{-0.2}$ & $0.5^{+0.3} _{-0.3}$ \\
Qatar-2b & $1360^{+60} _{-50}$ & $1100^{+200} _{-200}$ & $0.2^{+0.2} _{-0.2}$ & $0.6^{+0.2} _{-0.2}$ \\
WASP-43b & $1510^{+50} _{-50}$ & $650^{+90} _{-110}$ & $0.26^{+0.09} _{-0.1}$ & $0.09^{+0.07} _{-0.04}$ \\
WASP-52b & $1550^{+60} _{-50}$ & $1200^{+200} _{-200}$ & $-0.2^{+0.2} _{-0.3}$ & $0.6^{+0.2} _{-0.2}$ \\
WASP-140b & $1500^{+100} _{-100}$ & $1400^{+100} _{-100}$ & $-0.4^{+0.3} _{-0.4}$ & $0.9^{+0.2} _{-0.2}$ \\
Qatar-1b & $1540^{+60} _{-60}$ & $900^{+200} _{-200}$ & $0.3^{+0.1} _{-0.1}$ & $0.3^{+0.2} _{-0.1}$ \\
HD209458b & $1420^{+20} _{-20}$ & $1010^{+70} _{-80}$ & $0.52^{+0.04} _{-0.05}$ & $0.5^{+0.1} _{-0.1}$ \\
CoRoT-2b & $1760^{+40} _{-40}$ & $870^{+50} _{-40}$ & $0.33^{+0.06} _{-0.06}$ & $0.15^{+0.03} _{-0.03}$ \\
TrES-3b & $1700^{+100} _{-100}$ & $1000^{+200} _{-200}$ & $0.5^{+0.1} _{-0.2}$ & $0.3^{+0.3} _{-0.2}$ \\
WASP-77Ab & $1820^{+50} _{-50}$ & $1200^{+100} _{-100}$ & $0.35^{+0.07} _{-0.09}$ & $0.4^{+0.1} _{-0.1}$ \\
WASP-14b & $2560^{+60} _{-60}$ & $1730^{+60} _{-60}$ & $-0.8^{+0.1} _{-0.1}$ & $0.41^{+0.05} _{-0.05}$ \\
WASP-74b & $1750^{+60} _{-60}$ & $1270^{+90} _{-90}$ & $0.61^{+0.05} _{-0.05}$ & $0.5^{+0.1} _{-0.1}$ \\
KELT-14b & $2060^{+80} _{-70}$ & $1900^{+100} _{-100}$ & $0^{+0.1} _{-0.2}$ & $0.9^{+0.1} _{-0.1}$ \\
HAT-P-23b & $2290^{+100} _{-100}$ & $1100^{+200} _{-200}$ & $0.37^{+0.1} _{-0.11}$ & $0.13^{+0.1} _{-0.07}$ \\
WASP-19b & $2300^{+70} _{-70}$ & $1200^{+100} _{-100}$ & $0.38^{+0.07} _{-0.07}$ & $0.19^{+0.08} _{-0.06}$ \\
WASP-76b & $2580^{+60} _{-60}$ & $1400^{+200} _{-200}$ & $0.16^{+0.09} _{-0.1}$ & $0.2^{+0.1} _{-0.1}$ \\
KELT-20b & $2820^{+80} _{-80}$ & $700^{+200} _{-200}$ & $0.1^{+0.1} _{-0.1}$ & $0.01^{+0.01} _{-0.01}$ \\
WASP-121b & $2680^{+60} _{-60}$ & $1100^{+200} _{-200}$ & $0.31^{+0.06} _{-0.07}$ & $0.07^{+0.07} _{-0.04}$ \\
HAT-P-7b & $2900^{+100} _{-100}$ & $2500^{+200} _{-300}$ & $-0.7^{+0.3} _{-0.4}$ & $0.8^{+0.2} _{-0.2}$ \\
KELT-1b & $3200^{+100} _{-100}$ & $1500^{+200} _{-200}$ & $-0.2^{+0.2} _{-0.2}$ & $0.13^{+0.07} _{-0.05}$ \\
KELT-16b & $3100^{+100} _{-100}$ & $1200^{+400} _{-300}$ & $-0.1^{+0.2} _{-0.2}$ & $0.05^{+0.08} _{-0.04}$ \\
WASP-18b & $3130^{+50} _{-50}$ & $1160^{+80} _{-90}$ & $0.05^{+0.06} _{-0.07}$ & $0.05^{+0.02} _{-0.01}$ \\
WASP-103b & $2970^{+90} _{-90}$ & $900^{+100} _{-200}$ & $0.25^{+0.08} _{-0.09}$ & $0.02^{+0.02} _{-0.01}$ \\
WASP-12b (v1) & $2900^{+100} _{-100}$ & $1100^{+200} _{-200}$ & $0.36^{+0.09} _{-0.1}$ & $0.05^{+0.04} _{-0.03}$ \\
WASP-12b (v2) & $3000^{+100} _{-100}$ & $1500^{+200} _{-300}$ & $0.3^{+0.1} _{-0.1}$ & $0.2^{+0.1} _{-0.1}$ \\
MASCARA-1b & $3000^{+100} _{-100}$ & $1300^{+300} _{-300}$ & $0.32^{+0.09} _{-0.11}$ & $0.1^{+0.1} _{-0.1}$ \\
WASP-33b & $3230^{+50} _{-50}$ & $1560^{+40} _{-40}$ & $0.26^{+0.04} _{-0.04}$ & $0.13^{+0.01} _{-0.01}$ \\
KELT-9b & $4500^{+200} _{-200}$ & $3300^{+200} _{-200}$ & $0.2^{+0.1} _{-0.1}$ & $0.5^{+0.1} _{-0.1}$ \\
\enddata
\tablecomments{In the table, uncertainties are rounded to one significant figures and the parameter values are rounded to match the precision of their associated errors.}
\end{deluxetable} 

\begin{figure*}[t!]
	\includegraphics[width=\linewidth]{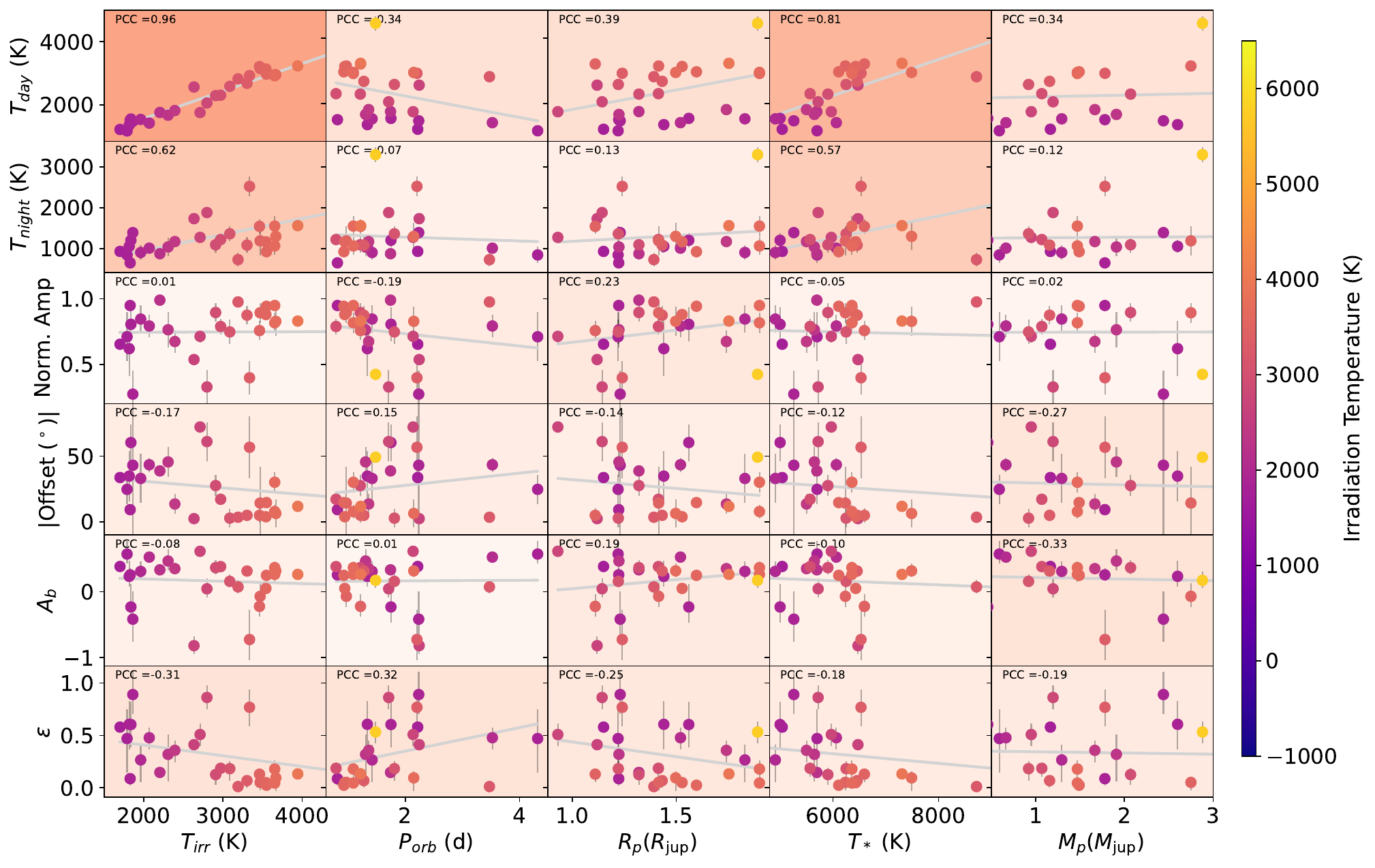}
 \vspace{-5mm}
    \caption{Trends in $T_{\rm day}$, $T_{\rm night}$, phase amplitudes and phase offsets against irradiation temperature, orbital period, planet's radius, stellar temperature and planet's mass. The color of the data point represent the irradiation temperature of the planet. The bestfit linear relationship between each pair of parameters is plotted in grey. The Pearson correlation coefficient (PCC) for each pair of independent parameter and measured quantity are printed on each panel and the background color of each panel scaled with the corresponding Pearson coefficient with the most opaque panel corresponding to a strong correlation. For irradiation temperature and planet masses, we have adjusted the axis limits in to focus on the densest parameter spaces as such the highly irradiated and massive planets do not appear on the plot.}
    \label{fig:trends_figure}
\end{figure*}

In the top two rows of Figure \ref{fig:trends_figure}, we confirm the positive correlation between dayside and nightside 4.5 $\mu$m brightness temperatures with irradiation temperature: planets that receive more flux from their star tend to be hotter. This is expected in the absence of secular trends in Bond albedo or day-to-night heat transport efficiency, as seen in multiple comparative Spitzer phase curve analyses \citep[e.g.,][]{2019NatAs...3.1092K,2021MNRAS.504.3316B,2022AJ....163..256M}. Closer inspection reveals a knee at irradiation temperatures of $\sim$2700 K, and a transition at this temperature has previously been reported in large surveys of Spitzer eclipse depths \citep{2020AJ....159..137G,2020A&A...639A..36B, 2023AJ....165..104D}. This knee has been attributed to a transition in the population caused by the onset of temperature inversions \citep{2020A&A...639A..36B}, changes in the efficiency of zonal circulation \citep{2023AJ....165..104D}, or the occurrence of cloud dissipation \citep{2023AJ....165..104D}.

The second row of Figure \ref{fig:trends_figure} shows that nightside 4.5 $\mu$m brightness temperature rises by only 50\% from 1000 K to 1500 K over more than a factor of two in irradiation temperature. As discussed by \cite{2021MNRAS.501...78P}, the weak dependence of nightside temperature with instellation could be a result of the $T^3$ dependence of the radiative timescale, i.e., hotter gas tends to cool faster, naturally leading to a nightside temperature that scales as $T_{\rm irr}^{1/3}$ or $T_{\rm irr}^{1/6}$, depending on whether wave-driven or jet-driven heat transport dominates, respectively. Additionally, the modest correlation is qualitatively consistent with silicate cloud coverage \citep{2019NatAs...3.1092K,2019AJ....158..166B}, while the slight upward trend is reminiscent of the gradual decrease in nightside clouds invoked by \cite{2021ApJ...908..101R}. 

There is also a trend between dayside and nightside temperature and stellar effective temperature, likely due to planetary demographics and observational bias. Ultra-hot Jupiters necessarily orbit hotter stars: a gas giant orbiting a cool star would be tidally disrupted before reaching irradiation temperatures of 3000 K. There could be hot Jupiters orbiting hot stars in longer orbits, but these would be expensive phase curve targets and hence are not in our sample. 

Three-dimensional models of hot Jupiters with different aerosol prescriptions predict different trends for the normalized phase amplitude as a function of irradiation temperature \citep{2021ApJ...908..101R}. In the absence of clouds, the normalized amplitude is expected to increase with irradiation temperature. In the presence of clouds, however, the amplitude would decrease with $T_{\rm irr}$. Regardless of the cloud prescription, they predict the normalized amplitude to be clustered around 0.7 at $T_{\rm irr} > 2700$ K. Our sample of hot Jupiters may include some cloudier planets and some with relatively clear skies, which would lead to greater scatter in normalized amplitude, even at a given irradiation temperature. Moreover, our sample of real-life hot Jupiters is undoubtedly more heterogeneous than the simulated planets of \cite{2021ApJ...908..101R}, which share similar rotation periods. Further curation of samples may be needed to tease out the expected trends in phase amplitude due to clouds.

The fourth row of Figure \ref{fig:trends_figure} shows the absolute value of the phase offsets. We expect this to be the hardest quantity to robustly extract from phase curves in the presence of detector drift or stellar variability. Moreover, \cite{2021MNRAS.504.3316B} reported that the derived phase offset can vary by over 10$^\circ$ depending on the detrending model used. Indeed, recent papers comparing the Spitzer/IRAC and JWST/NIRSpec phase curves of WASP-121b find large discrepancies between the Spitzer and JWST phase offsets \citep[cf.][]{2023A&A...676A..54M, 2023ApJ...943L..17M}. Nonetheless, we note that ultra-hot Jupiters generally exhibit smaller phase offsets, as one would expect from their shorter radiative timescales \citep{2011ApJ...729...54C} and stronger magnetic drag \citep{2022AJ....163...35B}. 

The fifth row suggests a lack of correlation in Bond albedo. However, in the region of Figure \ref{fig:BondAlb} where $A_{\rm B} > 0$, more irradiated planets also tend to have lower Bond albedos compared to cooler planets, which is consistent with the dissipation of dayside clouds on ultra-hot Jupiters \citep{2016ApJ...828...22P,2023AJ....165..104D}. Additionally, hotter planets tend to have lower day-to-night heat transport efficiency compared to their cooler counterparts. This is consistent with predictions from \cite{2011ApJ...729...54C} and \cite{2016ApJ...821...16K}: greater irradiation produces hotter atmospheres with shorter radiative timescales and hence poorer heat recirculation efficiency. Although one might expect the trend in normalized phase amplitudes as a function of irradiation temperature to mirror the trend in heat recirculation efficiency, the absence of such a trend could be attributed to datasets with suspiciously large phase offsets, and large $F_{\rm max}$. These phase curves may result from uncorrected systematics, which can significantly affect the measured maximum phase amplitude. It is also important to note that the circulation efficiency is computed in temperature space and does not account for the shape of the phase curve, making it more robust against uncorrected low-frequency systematics.

In rare cases (e.g., \citealt{2022AJ....163...32D}), a negative Bond albedo would indicate that the energy budget of the planet exceeds the level of irradiation received from the host star, e.g., due to tidal heating of eccentric planets. However, we have purposely excluded eccentric planets from our sample, so we expect these planets to have global energy budgets dominated by stellar irradiation. Indeed, the supposed negative albedos in our study tend to have large uncertainties, not to mention the systematic uncertainty in converting a brightness temperature to an effective temperature \citep{2019MNRAS.489..941P}, so we do not claim a detection of excess thermal emission.

Ultimately, the population of hot Jupiters is extremely diverse, with significant variations in parameters other than the ones we tested against. As demonstrated by \cite{2024MNRAS.531.1056R}, through a large grid of GCM models, the amplitude and offsets of phase curves can vary significantly at a given irradiation temperature due to other planetary factors such as metallicity, rotation, and gravity. For example, their models show that phase curve amplitudes can vary by as much as 0.4 and hotspot offsets can range from 0 to 30$^\circ$ for planets with the same levels of irradiation. To understand the intrinsic planet diversity through observed population-level trends, we must consider the inherent scatter created by differing planetary responses to these parameters. Careful curation of targets for future phase curve surveys will enhance our ability to better isolate and understand the effects of individual planetary parameters on atmospheric dynamics.

\section{Conclusion}\label{sec:conclusion}
We presented a comprehensive and uniform analysis of all Spitzer 4.5 $\mu$m phase curves of 29 hot Jupiters using the \texttt{SPCA} pipeline. BLISS mapping is the best detrending method for most of these datasets, but Spitzer systematics are not one-size-fits-all: other detector models are preferred for specific datasets. Regardless of which detrending scheme we adopt, we see the same trends, or lack thereof. Certain expected trends are robustly detected, but some expected trends are not manifest. We confirm that planets receiving greater instellation are hotter, but we are still far from confirming theories of heat transport in hot Jupiter atmospheres.

It will take larger sample sizes and higher-fidelity phase curve measurements from JWST and Ariel to firmly establish the system parameters governing day--night heat transport on synchronously rotating exoplanets. Possible avenues for improvement include: broader wavelength coverage to minimize the systematic error in going from emission observations to estimates of bolometric flux \citep{2019MNRAS.489..941P}, more uniform observing setups, better control of detector systematics, more planets, and better curation of target samples. As Figure \ref{fig:Spitzer sample} suggests, the thermal phase variations of hot Jupiters are complex beasts, and it will take much more work to tame them!

L.D.\ is a Banting Postdoctoral Fellow and would like to acknowledge funding from the Natural Sciences and Engineering Research Council of Canada (NSERC), and the Institut Trottier de recherche sur les exoplanètes (iREx). T.J.B.\ acknowledges funding support from the NASA Next Generation Space Telescope Flight Investigations program (now JWST) via WBS 411672.07.04.01.02. M.W.M.\ acknowledges support through NASA Hubble Fellowship grant HST-HF2-51485.001-A awarded by the Space Telescope Science Institute, which is operated by AURA, Inc., for NASA, under contract NAS5-26555. N.B.C.\ acknowledges support from an NSERC Discovery Grant, a Tier 2 Canada Research Chair, and an Arthur B.\ McDonald Fellowship, and thanks the Trottier Space Institute and l’Institut de recherche sur les exoplanètes for their financial support and dynamic intellectual environment.
\vspace{5mm}
\facilities{Spitzer(IRAC)}
\software{astropy \citep{2013A&A...558A..33A,2018AJ....156..123A}}

\appendix
\section{Target List In Context}\label{sec:Appendix Target Params}
Table \ref{tab:exoparams_table} shows the most precise constraints found on the exoplanet archives using the \texttt{exofile} pipeline\footnote{\url{https://github.com/AntoineDarveau/exofile}}. Using these parameters, we calculated the irradiation temperature $T_{\rm irr} = T_{\star} \sqrt{R_{\star}/a}$, the equilibrium temperature $T_{\rm eq} = (1/4)^{1/4} T_{\rm irr}$, and the surface gravity $g = GM_p/R_p^2$, where $G$ is the gravitational constant.
\begin{deluxetable*}{lCCCCCCCCl}[!h]
\tablecaption{Table of measured system parameters from the NASA Exoplanet Archives and derived quantities\label{tab:exoparams_table}}
\tablewidth{0pt}
\label{tab: params_orbital}
\tablehead{
\colhead{Planet} & \colhead{$T_{\star}$} & \colhead{$R_{\star}$ ($R_\odot$)} & \colhead{$P$ (d)} & \colhead{$a$ (AU)} & \colhead{$i$ ($^\circ$)} & \colhead{$R_p$ ($R_{j}$)} & \colhead{$M_p$ ($M_j$)} &  \colhead{$T_{\rm eq}$ (K)} & \colhead{$g$ ($m/s^2$)}}
\startdata
HD 189733 b & $5040^{+50} _{-50}$ & $0.76^{+0.02} _{-0.02}$ & $2.21857^{+3e-08} _{-3e-08}$ & $0.0311^{+0.0005} _{-0.0005}$ & $85.78^{+0.03} _{-0.03}$ & $1.12^{+0.04} _{-0.04}$ & $1.17^{+0.05} _{-0.05}$ & 1429 & 23.081\\ 
WASP-34 b & $5700^{+100} _{-100}$ & $1.11^{+0.06} _{-0.06}$ & $4.31768^{+5e-07} _{-5e-07}$ & $0.0524^{+0.0004} _{-0.0004}$ & $85.2^{+0.2} _{-0.2}$ & $1.2^{+0.1} _{-0.1}$ & $0.59^{+0.01} _{-0.01}$ & 1504 & 9.825\\ 
Qatar-2 b & $4640^{+50} _{-50}$ & $0.71^{+0.02} _{-0.02}$ & $1.33712^{+4e-08} _{-4e-08}$ & $0.0216^{+0.0006} _{-0.0006}$ & $89.7^{+0.5} _{-0.5}$ & $1.15^{+0.03} _{-0.03}$ & $2.6^{+0.9} _{-0.9}$ & 1527 & 48.73\\ 
WASP-43 b & $4120^{+100} _{-160}$ & $0.6^{+0.03} _{-0.04}$ & $0.813474^{+2e-08} _{-2e-08}$ & $0.0142^{+0.0004} _{-0.0004}$ & $82.1^{+0.1} _{-0.1}$ & $0.93^{+0.07} _{-0.09}$ & $1.8^{+0.1} _{-0.1}$ & 1537 & 51.012\\ 
WASP-52 b & $5000^{+100} _{-100}$ & $0.79^{+0.02} _{-0.02}$ & $1.74978^{+1e-07} _{-1e-07}$ & $0.0272^{+0.0003} _{-0.0003}$ & $85.4^{+0.2} _{-0.2}$ & $1.22^{+0.06} _{-0.06}$ & $0.46^{+0.02} _{-0.02}$ & 1545 & 7.623\\ 
WASP-140 b & $5300^{+100} _{-100}$ & $0.87^{+0.04} _{-0.04}$ & $2.23598^{+2e-07} _{-2e-07}$ & $0.0323^{+0.0005} _{-0.0005}$ & $83.3^{+0.5} _{-0.8}$ & $1.4^{+0.4} _{-0.2}$ & $2.44^{+0.07} _{-0.07}$ & 1565 & 29.166\\ 
Qatar-1 b & $4900^{+100} _{-100}$ & $0.8^{+0.05} _{-0.05}$ & $1.42002^{+4e-08} _{-4e-08}$ & $0.0233^{+0.0004} _{-0.0004}$ & $84.1^{+0.2} _{-0.2}$ & $1.14^{+0.03} _{-0.02}$ & $1.29^{+0.05} _{-0.05}$ & 1649 & 24.55\\ 
HD 209458 b & $6060^{+50} _{-50}$ & $1.16^{+0.01} _{-0.02}$ & $3.52475^{+3e-07} _{-3e-07}$ & $0.0463^{+0.0007} _{-0.0007}$ & $86.71^{+0.05} _{-0.05}$ & $1.32^{+0.02} _{-0.02}$ & $0.66^{+0.02} _{-0.02}$ & 1740 & 9.46\\ 
CoRoT-2 b & $5700^{+70} _{-70}$ & $0.9^{+0.02} _{-0.02}$ & $1.743^{+1e-07} _{-1e-07}$ & $0.028^{+0.0008} _{-0.0008}$ & $87.8^{+0.1} _{-0.1}$ & $1.41^{+0.06} _{-0.06}$ & $3.3^{+0.2} _{-0.2}$ & 1852 & 41.034\\ 
TrES-3 b & $5650^{+80} _{-80}$ & $0.81^{+0.01} _{-0.03}$ & $1.30619^{+3e-08} _{-3e-08}$ & $0.023^{+0.001} _{-0.001}$ & $82^{+0.3} _{-0.3}$ & $1.32^{+0.06} _{-0.06}$ & $1.91^{+0.08} _{-0.08}$ & 1939 & 27.171\\ 
WASP-77 A b & $5500^{+80} _{-80}$ & $0.95^{+0.01} _{-0.01}$ & $1.36003^{+7e-08} _{-7e-08}$ & $0.0234^{+0.0004} _{-0.0004}$ & $89.4^{+0.4} _{-0.7}$ & $1.23^{+0.03} _{-0.03}$ & $1.67^{+0.07} _{-0.06}$ & 2011 & 27.311\\ 
WASP-14 b & $6500^{+100} _{-100}$ & $1.3^{+0.1} _{-0.1}$ & $2.24377^{+2e-07} _{-2e-07}$ & $0.037^{+0.001} _{-0.001}$ & $84.6^{+0.2} _{-0.2}$ & $1.22^{+0.04} _{-0.04}$ & $7.8^{+0.5} _{-0.5}$ & 2214 & 129.017\\ 
WASP-74 b & $6000^{+100} _{-100}$ & $1.64^{+0.05} _{-0.05}$ & $2.13775^{+5e-07} _{-5e-07}$ & $0.037^{+0.001} _{-0.001}$ & $79.8^{+0.2} _{-0.2}$ & $1.56^{+0.06} _{-0.06}$ & $0.95^{+0.06} _{-0.06}$ & 2279 & 9.676\\ 
KELT-14 b & $5700^{+100} _{-100}$ & $1.52^{+0.03} _{-0.03}$ & $1.71005^{+1e-07} _{-1e-07}$ & $0.0296^{+0.0004} _{-0.0006}$ & $78.3^{+0.3} _{-0.3}$ & $1.5^{+0.1} _{-0.1}$ & $1.2^{+0.07} _{-0.07}$ & 2352 & 12.831\\ 
HAT-P-23 b & $5900^{+80} _{-80}$ & $1.2^{+0.07} _{-0.07}$ & $1.21289^{+7e-08} _{-7e-08}$ & $0.023^{+0.0003} _{-0.0003}$ & $85.7^{+0.9} _{-0.9}$ & $1.22^{+0.04} _{-0.04}$ & $2.1^{+0.1} _{-0.1}$ & 2444 & 34.247\\ 
WASP-19 b & $5570^{+70} _{-70}$ & $1^{+0.02} _{-0.02}$ & $0.788839^{+2e-08} _{-2e-08}$ & $0.0163^{+0.0002} _{-0.0002}$ & $78.8^{+0.6} _{-0.6}$ & $1.39^{+0.04} _{-0.04}$ & $1.07^{+0.04} _{-0.04}$ & 2497 & 13.675\\ 
WASP-76 b & $6200^{+100} _{-100}$ & $1.73^{+0.04} _{-0.04}$ & $1.80988^{+3e-07} _{-3e-07}$ & $0.033^{+0.0005} _{-0.0005}$ & $88^{+1} _{-2}$ & $1.83^{+0.06} _{-0.04}$ & $0.92^{+0.03} _{-0.03}$ & 2594 & 6.809\\ 
KELT-20 b & $8700^{+200} _{-300}$ & $1.56^{+0.06} _{-0.06}$ & $3.4741^{+2e-07} _{-2e-07}$ & $0.054^{+0.001} _{-0.002}$ & $86.1^{+0.3} _{-0.3}$ & $1.74^{+0.07} _{-0.07}$ & $3.382^{+0.0001} _{0}$ & 2682 & 27.656\\ 
WASP-121 b & $6500^{+100} _{-100}$ & $1.46^{+0.03} _{-0.03}$ & $1.27492^{+5e-08} _{-5e-08}$ & $0.026^{+0.0004} _{-0.0006}$ & $87.6^{+0.6} _{-0.6}$ & $1.75^{+0.04} _{-0.04}$ & $1.16^{+0.07} _{-0.07}$ & 2777 & 9.332\\ 
HAT-P-7 b & $6500^{+100} _{-100}$ & $1.99^{+0.07} _{-0.09}$ & $2.20474^{+2e-08} _{-2e-08}$ & $0.036^{+0.004} _{-0.002}$ & $82^{+1} _{-1}$ & $1.43^{+0.03} _{-0.03}$ & $1.78^{+0.08} _{-0.06}$ & 2804 & 21.527\\ 
KELT-1 b & $6600^{+200} _{-100}$ & $1.46^{+0.08} _{-0.07}$ & $1.21749^{+1e-07} _{-1e-07}$ & $0.0247^{+0.0002} _{-0.0002}$ & $86.8^{+0.8} _{-0.8}$ & $1.11^{+0.03} _{-0.02}$ & $27.2^{+0.5} _{-0.5}$ & 2910 & 547.794\\ 
KELT-16 b & $6240^{+50} _{-50}$ & $1.36^{+0.06} _{-0.05}$ & $0.968993^{+1e-07} _{-1e-07}$ & $0.0204^{+0.0002} _{-0.0003}$ & $84^{+3} _{-2}$ & $1.42^{+0.08} _{-0.07}$ & $2.8^{+0.2} _{-0.2}$ & 2916 & 34.044\\ 
WASP-18 b & $6430^{+50} _{-50}$ & $1.32^{+0.06} _{-0.06}$ & $0.941452^{+2e-08} _{-2e-08}$ & $0.0202^{+0.0003} _{-0.0003}$ & $84.9^{+0.3} _{-0.3}$ & $1.24^{+0.08} _{-0.08}$ & $10.2^{+0.4} _{-0.4}$ & 2978 & 164.427\\ 
WASP-103 b & $6100^{+200} _{-200}$ & $1.44^{+0.05} _{-0.03}$ & $0.925545^{+6e-08} _{-6e-08}$ & $0.0198^{+0.0002} _{-0.0002}$ & $88^{+2} _{-2}$ & $1.53^{+0.07} _{-0.05}$ & $1.49^{+0.09} _{-0.09}$ & 2984 & 15.818\\ 
WASP-12 b & $6400^{+100} _{-100}$ & $1.66^{+0.05} _{-0.04}$ & $1.09142^{+4e-08} _{-4e-08}$ & $0.0234^{+0.0006} _{-0.0005}$ & $83.4^{+0.7} _{-0.6}$ & $1.9^{+0.06} _{-0.06}$ & $1.47^{+0.08} _{-0.07}$ & 3071 & 10.093\\ 
MASCARA-1 b & $7500^{+200} _{-200}$ & $2.08^{+0.02} _{-0.02}$ & $2.14877^{+9e-07} _{-9e-07}$ & $0.04035^{+5e-05} _{-5e-05}$ & $88.4^{+0.2} _{-0.2}$ & $1.6^{+0.02} _{-0.02}$ & $3.7^{+0.9} _{-0.9}$ & 3083 & 35.959\\ 
WASP-33 b & $7300^{+100} _{-100}$ & $1.6^{+0.06} _{-0.05}$ & $1.21987^{+2e-07} _{-2e-07}$ & $0.0256^{+0.0002} _{-0.0002}$ & $87.9^{+0.9} _{-0.9}$ & $1.5^{+0.04} _{-0.04}$ & $3.3^{+0.7} _{-0.7}$ & 3316 & 36.278\\ 
KELT-9 b & $10200^{+400} _{-400}$ & $2.36^{+0.07} _{-0.06}$ & $1.48112^{+1e-07} _{-1e-07}$ & $0.035^{+0.001} _{-0.001}$ & $86.8^{+0.2} _{-0.2}$ & $1.89^{+0.06} _{-0.05}$ & $2.9^{+0.8} _{-0.8}$ & 4815 & 19.963\\ 
\enddata
\tablecomments{In the table, uncertainties are rounded to one significant figures and the parameter values are rounded to match the precision of their associated errors.}
\end{deluxetable*}

\section{Robustness of trends}\label{sec:robustness}
Since different model fits can sometimes yield different results, we experiment with different model selection prescriptions to ensure our reported trends are robust against our choice of model selection criteria. We also tested our statistical analysis presented in Figure 4 and found small variations in the resulting trends and dispersion when using different model selection criteria.

\begin{figure*}[b!]
    \centering
    \includegraphics[width=0.32\textwidth]{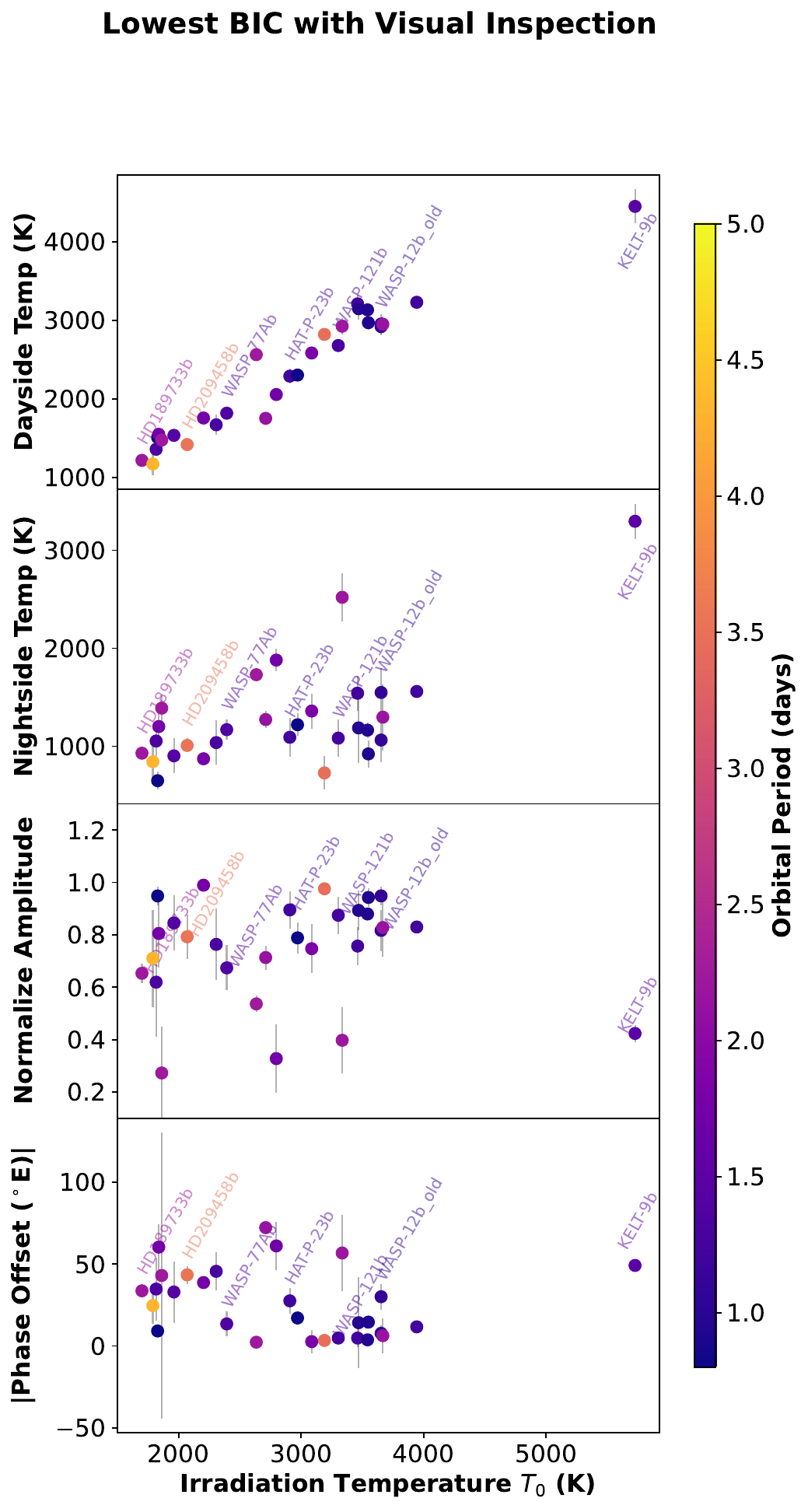}
    \includegraphics[width=0.32\textwidth]{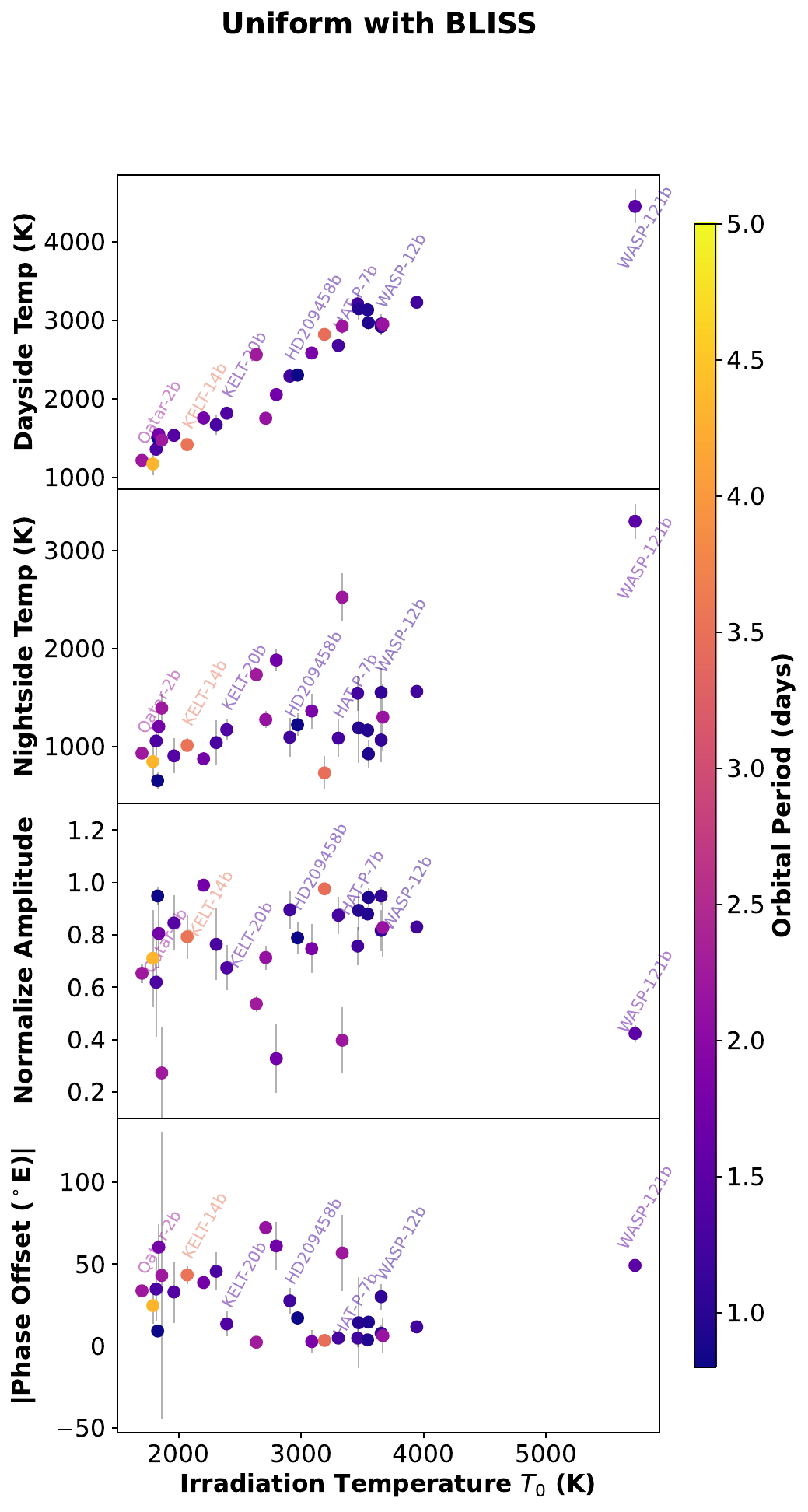}
    \includegraphics[width=0.32\textwidth]{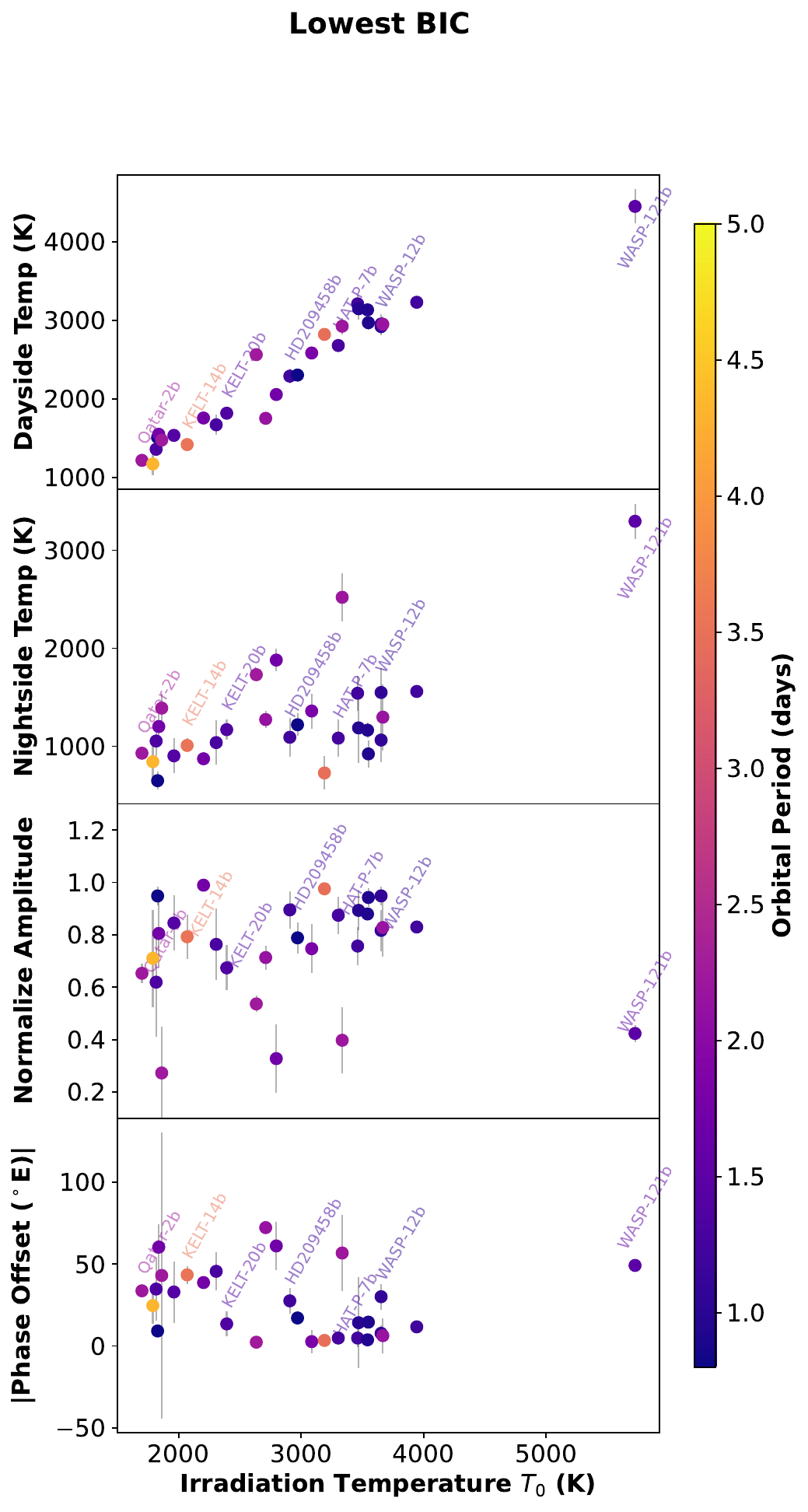}
    \includegraphics[width=0.32\textwidth]{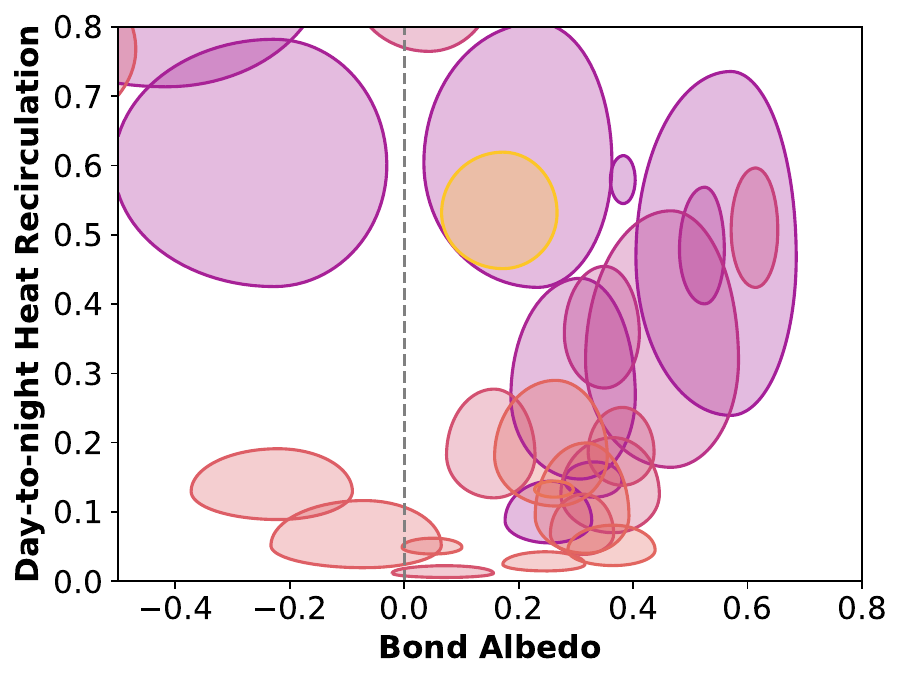}
    \includegraphics[width=0.32\textwidth]{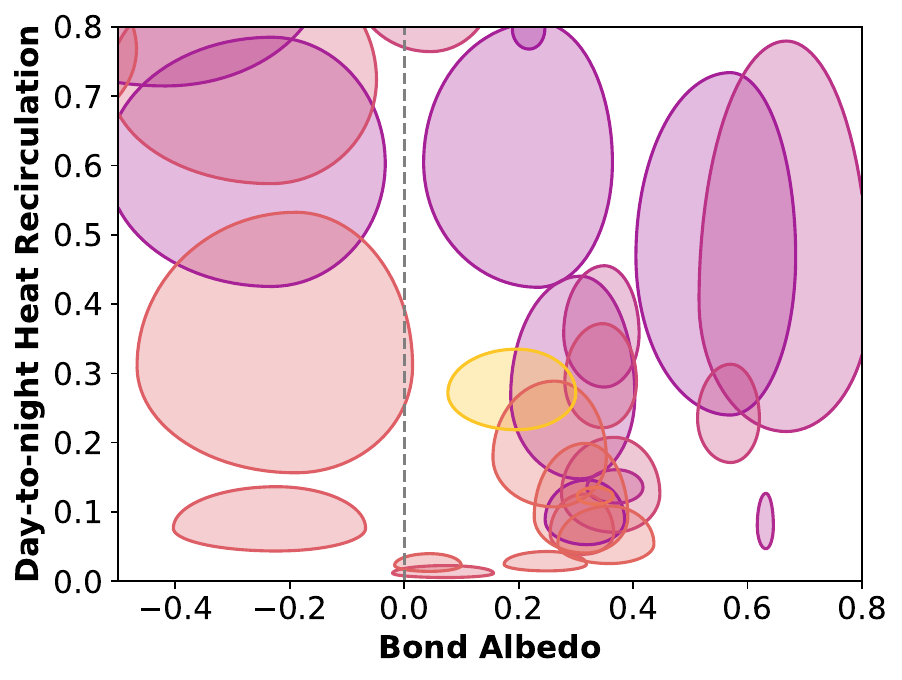}
    \includegraphics[width=0.32\textwidth]{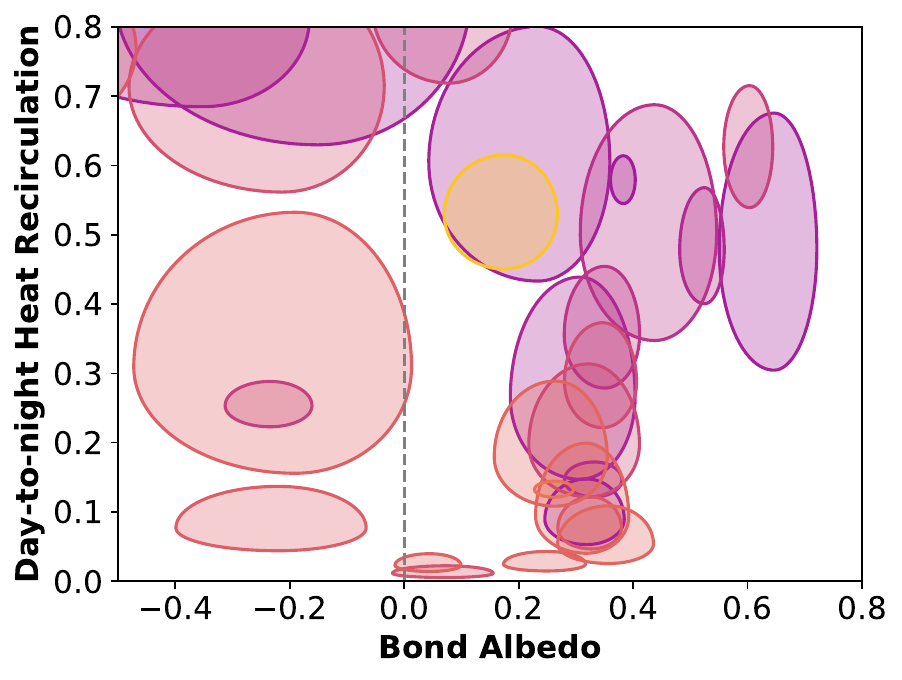}
    \caption{\emph{Top:} Trends in phase curves for different model selection strategies. The left plot is inferred using our nominal strategy with visual inspection, the middle plot is inferred from a uniform analysis using BLISS, and the right plot is calculated using the phase curve models preferred by the BIC. \emph{Bottom:} Trends in Bond albedo and Heat Recirculation Efficiency for different model selection strategies. The left plot is inferred using our nominal strategy with visual inspection, the middle plot is inferred from a uniform analysis using BLISS, and the right plot is calculated using the phase curve models preferred by the BIC.}
    \label{fig:robustness_trends}
\end{figure*}

\bibliography{UltimatePC}{}
\bibliographystyle{aasjournal}
\end{document}